*Review*

# Research Needs for Realization of Zero-Carbon Power Grids with Selected Case Studies

**Young-Jin Kim**

Department of Electrical Engineering, Pohang University of Science and Technology (POSTECH), Pohang, Gyungbuk 37673, Korea

**Abstract:** The attainment of carbon neutrality requires a research agenda that addresses the technical and economic challenges that will be encountered as we progress toward 100% renewable electricity generation. Increasing proportions of variable renewable energy (VRE) sources (such as wind turbines and photovoltaic systems) render the supply-and-demand balance of VRE-dominated power grids difficult. The operational characteristics and effects of VRE inverters also require attention. Here, we examine the implications of the paradigm shift to carbon neutrality and summarize the associated research challenges in terms of system planning, operation, and stability, and the need for energy storage integration, demand-side participation, distributed control and estimation, and energy sector coupling. We also highlight the existing literature gaps, and our recent studies that can fill in the gaps, thereby facilitating the improvement of grid operation and estimation. The numerical results of comparative case studies are also provided on the operational stability and economics of power grids with a high level of VRE sources, assisting stakeholders in establishing specific roadmaps and making relevant decisions.

## 1. Introduction

Worldwide decarbonization of electric power grids is essential to halt climate change caused by greenhouse gas emissions [1,2]. A deep commitment to 100% renewable electricity poses challenges in terms of power grid operation, control, and protection; we must overcome these challenges in a cost-effective manner. The no-regrets pathway toward carbon neutrality requires analysis of the principal challenges that will be encountered when incorporating renewable energy into current power grids and how these can be solved via further deployment of renewable energy sources and new energy-related technologies [3]. In other words, the no-regrets pathway does not begin in a barren landscape, rather in a place surrounded by existing grid assets, the use of which must be optimized. A 100% penetration of renewable energy is an endpoint of a long path where all different penetrations are assured, long term [4].

### 1.1. Real-World Pathways

Since the Paris Agreement was adopted in 2015, nations worldwide have made major efforts to implement decarbonized power grids. For example, in the United States, over 200 cities and counties plan to use only renewable energy to meet all of their electricity needs over the next decades [5]. Four states have already enacted 100% clean energy mandates, and a further thirteen are drafting similar legislation [6]. Six more states have introduced nonbinding goals of 100% renewable electricity [4]. South Korea has recently finalized a plan to achieve carbon neutrality by 2050 [7–9], mainly by restricting coal and natural gas power plants and replacing internal-combustion-engine vehicles (ICEVs) with electric vehicles (EVs) and hydrogen-powered vehicles (HPVs). Two scenarios were considered when evaluating the sectors of electricity, transportation, and hydrogen, for example. The details differ; flexible adaptive measures can be taken when unexpected environmental and business events arise. Specifically, in the first scenario, all thermal power plants using coal, natural gas, and oil are scrapped. The second scenario abolishes coal-fired power plants but retains natural gas-fired plants to ensure grid reliability, creating 20.7 million metric tons of carbon dioxide in 2050 (92.3% less than the 269.6 million metric tons of 2018). In the second scenario, the transformation of coal-fired plants to natural gas-fired plants is considered; carbon-capture and storage (CCS) technologies will be deployed to compensate for the carbon emissions. The proportion of renewable energy is planned to increase from 6.2% in 2018 to 30.2% in 2030, while the share of nuclear energy will increase marginally to 23.9% in 2030 from 23.4% in 2018. In terms of the transportation sector, the scenarios require that 97% and 85% (respectively) of all vehicles are EVs or HPVs. The remaining ICEVs use environmentally friendly fuels and are fitted with direct-air-capture devices. It is envisaged that the transportation sector will respectively emit up to 2.8 and 9.2 million metric tons of carbon dioxide in 2050, thus 97.1% and 90.6% less than the 98.1 million metric tons of 2018. Other countries are planning along similar lines [10], although small grid-specific differences are apparent.

### 1.2. Definition and Research Needs

When focusing on carbon neutrality on a national scale, it is first necessary to define a zero-carbon power grid. Renewable energy sources can be categorized into two types [11]: weather-dependent sources, such as wind turbines and photovoltaic (PV) arrays, often termed variable renewable energy (VRE) sources, and weather-independent sources including nuclear, hydroelectric, biomass, and geothermal power plants. The first- and second-type sources use power electronics converters and synchronous machines, respectively, to convert non-electric energy to electric energy [12]. A zero-carbon power grid is not simply a power grid including only VRE sources (VRE share of total power generation is 100%). For example, in 2019, the United States acquired approximately 10% of its electricity from non-VRE sources; this is expected to grow by 2050 [13]. The current power grids of Quebec, Norway, Iceland, and Austria rely heavily on hydroelectric, biomass, geothermal, and solar power, and already operate at or very close to 100% renewability [14,15]. However, the implementation of zero-carbon power grids using VRE sources, primarily wind power and solar PV, is receiving greater attention, given their global abundance and the lack of geographical constraints. In other words, VRE will account for a large proportion of future power. Rapid VRE integration will be the primary pathway toward zero-carbon grids worldwide [4].

An understanding of the challenges posed when moving toward a zero-carbon grid is essential when identifying and developing a research agenda that lays out what is and is not known, and the first and the best solutions. The understanding will enable the establishment of a research roadmap, assisting participants in prioritizing and focusing effort and investment to achieve the right solutions of many issues associated with zero-carbon power grids in a reliable and sustainable manner. For example, a power grid with instantaneous VRE penetration of 100% will take considerably shorter to achieve that of a grid with an annual penetration of less than 80% [15,16]. This is principally because existing technologies can assist penetration to 80% but the final 20% requires new research. System operators then need to understand what challenges will become critical during power grid decarbonization and have a plan on how and when to resolve them.

#### 1.2.1. Technologies of Interest

Every power grid differs technically, socially, and economically; different pathways toward decarbonization are required. Most grids with high proportions of VRE will undergo intensive renovation in terms of energy storage, sector coupling, demand response (DR), distributed control and estimation, and power electronics installation [2,4,14,15,17]; these are not required by grids with abundant renewable energy that is less variable. Although the technologies are very powerful, their synergistic contributions are yet to be revealed; it is essential that they be well integrated to ensure reliable and cost-effective grid decarbonization.

Recently, battery costs have declined dramatically, facilitating VRE use in short-term (e.g., daily) grid operations. Battery storage is unlikely to be viable in longer term (e.g., seasonal) operations [18]. Hydrogen tanks can be a good alternative for seasonal storage [19], although power-to-gas technologies remain very costly, particularly when operating in the classical electricity-in/electricity-out modes; however, the costs are decreasing. Thermal energy storage also can meet longer-term storage needs [20].

In power grids, the demand side will be subject to major changes as the economy and industry continue to be electrified. Existing loads can be better managed using communication and command-and-control systems, or can be replaced by new loads evidencing better operational flexibility. The demand-side changes will include power-to-X conversion and further energy sector coupling, whereby electricity systems play central roles in large-scale coupling to thermal, natural gas, and hydrogen systems. Load coupling will provide more options to compensate for VRE intermittency and inflexibility. The social transition to digitalization will facilitate demand-side changes, further increasing demand-side participation in future power grids [21].

Moreover, the sites of VRE production will become closer to the locations of energy consumption. This implies the need for a distributed paradigm of power grid control and estimation, rather than a centralized paradigm, whereby bulk power is delivered to remote load centers in a unidirectional fashion via high-voltage, long transmission lines. In regions with significant levels of distributed VRE sources, distribution management systems (DMSs) require special attention; these have not traditionally reflected the sources of active generation and the controllable loads [22]. When VRE sources are aggregated across larger distribution areas using high-voltage transmission lines, a hybrid approach that combines the best of the centralized and distributed paradigms will be required in practice for the realization of 100% renewable electricity.

Another essential technology is that of power electronics converters [2,4,14,15], by which many VRE sources are interfaced at points of common coupling (PCCs) spread across power networks. Power electronics converters for VRE sources differ from the traditional synchronous machines used by fossil fuel generators, with respect to synchronization, programmability, dynamic responses, power quality, and protection. Therefore, the siting, control methods, and operating characteristics of VRE inverters significantly affect power grid stability and reliability and, further, transmission line expansion and even electricity markets and regulations.

### *1.3. Paper Objective, Scope, and Structure*

Zero-carbon power grids can be realized by solving many research challenges that differ with the time and resources required, although the technological pathways remain somewhat uncertain because multiple paradigm shifts will impact grid planning, opera-tional scheduling, and real-time control. This paper describes the main research challenges to be solved when moving to 100% renewable electricity and presents relevant case studies. There is an urgent need for research on supply-and-demand balancing and incorporation of large numbers of VRE inverters into grids. The case studies offer new solutions for improved grid operation, estimation, and protection. They use state-of-the-art power and control technologies. They are discussed with numerical results from the perspectives of originality, feasibility, and effectiveness. We expect that this will attract the attention of universities, research institutes, and funders, and encourage the establishment of roadmaps or plans that prioritize effort and investment. This paper focuses on the technical and economic aspects of the challenges and feasible, implementable solutions; we do not address policy. Although this paper focuses on transmission and distribution networks in the United States and Europe, most of the research challenges and solutions apply more generally. The trends in zero-carbon power-grid establishment, and the difficulties encountered, are similar in other countries.

A few surveys on zero-carbon power grids were recently conducted, for example, in [1,2,4,24]. It is the opinion of the authors that this paper complements the existing surveys in that we (1) comprehensively searched for the recent literature, most of which were published from 2016 to 2022 (apart from a few original works); (2) selected the literature mainly in four areas that significantly affect the realization of zero-carbon power grids (i.e., supply-and-demand balancing, inverters as VRE interfaces, energy and capacity adequacy, and electricity market design); (3) further classified the four main areas into 15 sub-areas; (4) outlined challenges and future research directions for each sub-area; and (5) presented the results of comparative case studies selected from the literature that are relevant with more than one challenge simultaneously.

The remainder of this paper is organized as follows: Section 2 presents the critical research agenda and the specific challenges, Section 3 discusses relevant case studies and their numerical results, Section 4 describes the further research required, and Section 5 concludes the paper.

## 2. Critical Research Agenda

A power grid must reliably and cost-effectively balance power supply and demand at all locations and at all times [23,24]. VRE generation is weather dependent and thus difficult to predict, unlike power generation from thermal, nuclear, and hydroelectric plants. Moreover, VRE sources are connected to electric grids via power electronics converters; these differ from synchronous machine-based generators.

### *2.1. Supply-and-Demand Balancing*

When designing a zero-carbon power grid, it is essential to define the system requirements and communicate a consistent set of well-defined balancing services [2,12] to all system participants across the spatial and temporal domains; this ensures the efficient and reliable operation of current and future grids.

#### 2.1.1. Timeframes

Traditionally, supply-and-demand power balancing is a major operational challenge; energy and capacity adequacies were addressed in the planning stage, as discussed in Section 4.1. However, in a zero-carbon power grid, wind and solar PV energy vary significantly, increasing the extent and duration of supply-and-demand imbalance compared to those of a traditional grid producing weather-independent energy. Thus, apart from energy provision, a zero-carbon grid must feature the flexibility and reliability required to maintain the supply-and-demand balance at the desired level from the planning stage to the scheduling, and to the real-time operation [24]. In other words, as we move toward 100% renewable electricity, power balance and system adequacy must be simultaneously tackled in the planning and operational stages, and on both the supply and demand sides.

#### 2.1.2. Forecasting

Given VRE uncertainty, forecasting must improve for the optimal use of balancing resources. VRE forecasting errors increase reserve procurement and render the commitment of generators and storage units sub-optimal. Forecasting of wind and solar PV energy must improve along with forecasting of the weather, such as wind speed, solar irradiance, and ambient temperature [25]; weather forecast uncertainties seriously affect when and how wind and PV generators cannot only supply energy but also provide essential services for dynamic ramping, a fast-operating reserve, and the mitigation of line congestion [26–28], for example. Moreover, forecasting must integrate weather-dependent supply (i.e., VRE sources) with weather-driven demand (e.g., electrified transportation and space heating and cooling) [29]. In other words, forecasting must consider all uncertainties in power generation and transmission and weather-driven demand together to capture the correlations thereof. This requires a shift from deterministic forecasting to probabilistic, stochastic forecasting. It is increasingly important to forecast not only power supply, delivery, and demand, but also their uncertainties [29]. For example, what will be the moment of inertia at time $h$? By how much will the frequency regulation reserve capacity and dynamic line rating fall at time $h$? How much of the total load demand will be flexible at time $h$? Such forecasting outputs can be incorporated into algorithms and tools used for optimal scheduling, outage assessment, and real-time control.

#### 2.1.3. Algorithms and Tools

The algorithms and technical tools of traditional power grids must be improved to ensure a stable balance between supply and demand in power grids with high portions of VRE. Power balancing then becomes more dynamic and uncertain, requiring a change from deterministic algorithms and tools to more probabilistic methods [27]. This aids grid operators to maximize the availability of existing flexible resources and hence minimize the reinforcement of operational flexibility. As we move toward 100% renewable electricity, emerging flexible resources, such as EVs and heating, ventilation, and air-conditioning (HVAC) units, enhance the flexibility of grid operations. To encourage customer engagement, the prices and incentives offered by the electricity, transportation, and heating/cooling sectors must be both transparent and practical, thus based on accurate assessments of the contributions of flexible resources. As such contributions have expanded on sector coupling, algorithms and tools that handle data acquisition, pre-processing, and visualization will become increasingly important and complex. It will become more difficult to remain fully aware of all grid operating states, and to maintain the supply-and-demand balance at all times. Moreover, the large increases in the numbers of flexible resources within distribution networks render it essential to develop interfacing algorithms and tools that estimate the availability, duration, and responsiveness of grid services for transmission networks [2,15]. This will facilitate coordination across remotely located regional grid operators at different voltage levels, fulfilling their new roles in system operations and interconnections.

#### 2.1.4. Modeling and Analysis

An increased reliance on VRE is likely to create more bottlenecks, which is presently the case within transmission and distribution networks. In the absence of solutions, this would curtail VRE use and cause failure of supply-and-demand balancing. Various technologies of power flow control mitigate the immediate need for reinforcement of transmission and distribution networks. These include dynamic line rating (DLR), flexible AC/DC transmission systems, and high-temperature low-sag (HTLS) conductors [30–32]. This implies that network modeling and power flow analysis should be improved to estimate and minimize potential curtailments of VRE use across networks with different voltages [33]. Unit commitment and dispatch models and grid expansion models must also be improved to fully exploit the new power flow control technologies and analyze their individual and synergistic effects. Moreover, future grid expansion modeling and analysis tools should be continually expanded to include the existing and forthcoming gas and hydrogen infrastructures and the up-to-date power-to-X conversion technologies [34].

Traditionally, power grid models were often simplified to facilitate the analysis of large-scale networks with many components while still ensuring high analytical accuracy. However, a very detailed model of a grid with 100% renewable electricity is required when analyzing the uncertainty and variability of supply-and-demand balancing and, consequently, the stability of grid operation, as discussed in Section 2.2. Moving from hour-resolution modeling to minute- or second-resolution modeling will greatly improve understanding of the real-time power balancing and dynamic responses of future grids with large numbers of VRE inverters. For example, unit commitment and dispatch models are widely used to assess the hourly or 5–15 min operations of generators. When updated with stochastic decision-making schemes of higher resolutions due to better VRE forecasting, the models will better assess the operational flexibility of generators and allocate operating and regulation reserves over a rolling planning horizon. The commitment and dispatch models can be further improved by having them reflect real-time power balancing issues, such as inertial responses, and frequency and voltage controls, under normal and abnormal grid conditions.

#### 2.1.5. Interactions across Regions and Sectors

Interactions among and control across geographical or administrative regions are important in terms of the overall supply–demand balance of power grids with high proportions of VRE because complementary flexible resources may well be available in neighboring regions [35,36]. Smart grid ICT facilitates optimal coordinated control of flexible resources located anywhere in the grid. Moreover, optimal interaction across different energy sectors (e.g., transportation, natural gas, and heating and cooling) within the same and other regions can be achieved via unit commitment and dispatch that reflects the process-specific constraints on each energy sector [2,37], e.g., EV driving patterns, seasonal space heat usage, and industrial heating. This will afford high supply- and demand-side flexibility to regional grid operators, further increasing the proportion of VRE. This is because sector coupling can be implemented not only on the supply side, but also on the storage and demand sides, opening alternative energy delivery paths and uses in different sectors.

#### 2.1.6. A List of Research Challenges

In terms of the sub-areas discussed above, the following challenges, listed in Table 1, are apparent when seeking to improve grid operational flexibility and thus reliable and cost-effective balancing of supply and demand.

**Table 1.** List of research challenges in balancing supply and demand.

| Challenges | Research Questions and Requirements |
|---|---|
| Timeframes | • Services that ensure a stable supply-and-demand balance and flexible power delivery should be expanded and ***incorporated at different timeframes.*** |
| Forecasting | • ***Long-term datasets with high spatiotemporal resolution*** should be collected, processed, and utilized to ensure accurate forecasting of VRE generation patterns.<br>• ***High-quality assessment of forecast uncertainty*** is required to integrate the weather-dependent characteristics of VRE sources into the operational algorithms and analytical tools of grids. |
| Algorithms and tools | • VRE curtailment caused by network congestion can be avoided by ***improving the accuracy of power flow analyses by enhancing network modeling and nonlinear solvers.***<br>• It is necessary to capture more details for each service to ***aggregate the technical details of constraints on individual resources into a single model or algorithm.*** |
| Modeling and analysis | • ***Stability constraints*** should also be represented better, including the limits of stored rotational energy, the frequency regulation reserve, and locational voltage deviations.<br>• To establish risk-aware power balancing, new optimization methods and new computations should be used ***to update the deterministic models and tools to the stochastic status***, and to further develop existing probabilistic models and tools. |
| Interactions across regions and sectors | • Energy storage systems and controllable loads in low-voltage networks are cost-effective resources for providing essential reliability services to bulk transmission systems. This requires ***a detailed representation of the complex constraints pertaining to aggregation of distributed energy storage and loads.***<br>• ***Coupling of energy sectors must be modeled in sufficient detail (i.e., high resolution) in terms of both flexibility and process-specific constraints***, making it possible to include many flexible resources in decarbonized power grids and greatly improve grid operations and economics. |

### *2.2. Inverters as VRE Interfaces with Grids*

The technological differences between power electronics converters and synchronous generators mean that future power grids with high proportions of VRE require fundamental changes in their modeling, analysis, control, and protection schemes [38–40]. Modeling tools for inertia-less grids require more detail and a higher time resolution than those available today. The real-time stability and controllability of different types of VER inverters (e.g., grid-forming, -supporting, and -following inverters) need to be analyzed more accurately for grids with high shares of VRE. New approaches toward frequency and voltage regulation; black-start capability; and distributed control of generation, storage, and load assets are required. Traditional protection schemes for power grids with high proportions of synchronous generators are subject to significant modification given the different characteristics of the short-circuit currents in VRE-dominated power grids.

#### 2.2.1. Desynchronization

Increasing proportions of VRE will change the underlying nature of a grid from synchronous to non-synchronous. The challenges in the use of VRE include a precise understanding of existing VRE inverters and their effects on the behavior of a VRE-dominated grid. Specifically, wind and solar PV generators are non-synchronously connected to the network via power electronics converters [38,40]; this stands in contrast to synchronous generators. The converters decouple the inertial responses of rotating machines (wind turbines) from the network. Additionally, solar PV lacks an inherent inertial response, reducing the moments of inertia and thus increasing the rate of frequency change. Moreover, compared to synchronous machines, power converters have a limited capacity to supply transient currents during grid-side faults [41]; this means that protection relays and schemes must change. The decoupled (or near-absent) inertial responses and the limited fault currents exert major effects, for example, on grid frequency and voltage regulation, the dynamic responses during faults, black-start capability, and power quality, such as current harmonics. However, power electronics converters are eminently programmable; they can emulate a variety of dynamic behaviors, in contrast to the unprogrammable nature of synchronous generators. This opens the possibility to deliver 100% renewable electricity in a reliable and cost-effective manner.

#### 2.2.2. Modeling and Analysis

Increasing attention is being directed to integration of VRE inverters and their controllers into positive-sequence, fundamental-frequency simulation tools. However, the development and analysis of fast inner-loop controllers are challenging, both to simulate and explore experimentally. Simulation requires faster time steps and more complex functionalities to represent the non-linear dynamics of VRE sources and the interfacing inverters [2,4,42]. For example, the phase-locked loop (PLL) controllers of VRE inverters must be accurately and completely modelled to foresee whether a grid interfacing with the controllers, containing few synchronous generators and thus with low system strength, will survive large disturbances and, if not, when the grid becomes unstable [42]. However, in existing simulation tools, the PLL controller models are greatly simplified. Such tools cannot model future power grids effectively. New modeling and analysis issues are also apparent, caused by behaviors at non-fundamental frequencies and control interference attributable to sub-synchronous resonance [43]. This further weakens the utilities of existing positive-sequence fundamental-frequency simulations tools. Recently, electro-magnetic transient (EMT) tools with fast time steps and more comprehensive models have received extensive attention [44]. However, EMT tools cannot be used for real-time modeling or analysis of an electrical network with more than hundreds of components because the computation time becomes excessive. To overcome this, co-simulation using positive-sequence and EMT tools is increasingly considered to be promising to balance modeling accuracy with computational efficiency.

#### 2.2.3. Frequency and Voltage Stabilities

Power systems have been analyzed using rather simple representations of cycles-to-seconds timeframe dynamics. Specifically, quasi-steady-state analyses have been widely used to assess both short-term dynamics and longer-term electro-mechanical dynamics, rather than real-time analyses. However, the greater the VRE proportion, the greater the supply-side variability and uncertainty, increasing the number of real-time challenges to the paramount role played by the system operator in terms of grid frequency and voltage regulation.

In most power grids, VRE already affects the frequency and voltage stability, which can, at least to some extent, be addressed using existing technologies and mitigated through conventional measures [45]. For example, low inertia and poor frequency stability have been addressed by reducing the minimum generated outputs and installing more synchronous condensers. Note that synchronous generators can remain connected in the form of synchronous condensers [4]. Similarly, all static synchronous compensators (STATCOMs), static VAR compensators (SVCs), synchronous condensers, and fixed and switchable shunt capacitors can support voltage stability. Moreover, rapid frequency and voltage controls for VRE sources are under vigorous development, particularly, for variable-speed wind turbines. Such mitigation measures will continue to evolve as VRE proportions increase. However, the upcoming high proportions of VRE call for more detailed models, better tools for stability assessment, and new ways to control frequency and voltage. This is principally because the stability and control principles for a VRE-dominated grid remain poorly understood. For example, it is not yet clear whether grid frequency is the best, right indicator of the supply-and-demand balance [38]. The proliferation of VRE inverters promises a seamless transition from existing AC-based transmission systems to DC-based systems of constant frequency (i.e., zero). DC applications are slowly but steadily growing in the form of high-voltage DC (HVDC) transmission systems and medium- and lower-voltage DC (MVDC and LVDC) distribution systems [46]. Fundamental decisions should then include, for example, the selection of suitable indicating parameters in the AC/DC hybrid grid, the role of frequency relative to the rotational speed of a synchronous machine, and the sensitivity of frequency-dependent loading.

#### 2.2.4. Short-Circuit Currents

Short-circuit current magnitudes over time are used to estimate grid impedance and hence, network strength [41]. The replacement of synchronous machines (traditional generators) by the power electronics converters of VRE sources significantly reduces the magnitudes of short-circuit currents, triggering unexpected unstable interactions between the operation and control of nearby VRE sources and other inverter-interfaced devices on the supply, storage, and demand sides. Moreover, individual VRE inverters exhibit different responses to grid-side faults. This renders it complicated in practice to calculate the short-circuit impedances of individual VRE inverters, particularly when unbalanced faults are in play, and hence, establish the appropriate schemes of protection coordination and fault-ride-through (FRT) operation of VRE-dominated grids [47]. Most existing positive-sequence simulators lack the detail required for accurate short-circuit analysis of wind turbines and solar PV generators. The simulators simply assume that VRE inverters operate only in a grid-following mode, rather than in a grid-forming or grid-supporting mode, underlining the concerns on the performance of VRE inverter controllers during faults and the effects thereof on voltage stability and controllability in a VRE-dominated grid. To address these issues, EMT-level simulators with high fidelity of models and controllers are essential to improve analysis of short-circuit currents and establish new protection and FRT schemes with high performance.

#### 2.2.5. Behind-the-Meter Units

Although most of the focus is on VRE inverters, end-user loads must be similarly considered because power electronics converters are also widely used as the interfaces of building loads, such as light-emitting diodes, EV chargers, and HVAC units [48,49]. Under converter control, the penetration of constant-power loads has increased sharply in the past decades, affecting the damping ratios of the interfacing power grids. Therefore, more detailed modeling of building loads is required to analyze their effects on the dynamic responses of the interfacing grids comprehensively. Moreover, the increasing numbers of power electronics converters increase the complexity of load dynamic behaviors, particularly over very short timeframes. In other words, loads can no longer be treated using simple constant-power, -current, or -impedance assumptions during EMT-level modeling and analysis. Another important factor is the emergence of behind-the-meter generation and storage units, which are also interfaced using power electronics converters [48], [49]. Such units significantly alter the characteristics of buildings' net load demands.

As the numbers of behind-the-meter units continue to increase, improved modeling and analysis become important, facilitated by advanced metering infrastructure and Internet-of-Thing (IoT) sensing technologies.

2.2.6. A List of Research Challenges

The establishment of future power grids with high proportions of VRE poses specific challenges (see Table 2) in the sub-areas discussed above. These challenges apply at both the transmission and distribution levels. The latter poses additional challenges in terms of decentralization.

**Table 2.** List of research challenges in VRE inverter interfacing.

| Challenges | Research Questions and Requirements |
|---|---|
| Desynchronization | • How can the change from a synchronous to a non-synchronous system be ***seamless while maintaining reliability?***<br>• Existing synchronous generators and network infrastructures should become ***more adaptive and co-operative to meet the increased flexibility, stability, and control needs of VRE-dominated grids***. |
| | • How can ***inverter design and the functionality of VRE-dominated power grids be optimized***?<br>• It is necessary to ***implement various forms of VRE inverter control*** that reference local conditions, parameter settings, and controller tuning, and to ***ensure that these are incorporated with grid-level control***. |
| Modeling and analysis | • It is necessary to ***understand the limitations of existing simulation models and tools for power electronic converters*** and to ***develop new models and tools that support the planning and interconnection of VRE-dominated grids***.<br>• ***Existing, positive-sequence fundamental-frequency simulation tools must be updated*** to represent faster controllers with advanced functionality and limiting conditions, focusing on accurate VRE inverter controller representation.<br>• ***Analytical simulation models*** that are needed to handle the characteristics of inverter-based generators, storage, and loads with real-world, complex, non-linear operational characteristics.<br>• ***The fidelity of generic and manufacturer-specific EMT-level simulation models and tools*** for large-scale network studies must be enhanced. |
| Frequency and voltage stabilities | • ***Which parameters best indicate the supply-and-demand balance and how the grid frequency should be controlled?***<br>• It is necessary to ***understand the effects of system harmonics and sub-synchronous oscillations from VRE inverters on grids and to design mitigation method,*** such as shaping of the inverter harmonic impedances at specific frequencies. |
| Short-circuit currents | • Traditional protection schemes, designed for synchronous generators should be replaced by ***new protection schemes for a wide variety of VRE inverters.***<br>• New methods are needed for the ***restoration and black start of VRE-dominated grids.*** |
| Behind-the-meter units | • Detailed planning and operating applications are needed to ***integrate various DC devices into an AC power grid***. |

## 3. Selected Case Studies and Numerical Results

In this section, case studies that address the research challenges discussed in Section 2 are presented. With comparative, numerical results, the cases illustrate new algorithms and strategies that improve grid operation, estimation, and protection from the perspectives of originality, feasibility, and effectiveness.

### *3.1. HVDC Applications and Optimal Control*

Offshore wind farms (OWFs) have attracted significant attention; these reduce fossil-fuel emissions and mitigate climate change [50]. HVDC technology is a promising means by which remote, large-scale OWFs can be connected to a main power grid. HVDC technologies include both line-commutated converters (LCCs) and voltage source converters (VSCs); these differ. Specifically, LCC control strategies are mature and trusted. An LCC can transmit bulk power with little loss, and it is inexpensive to install [51]. Several ongoing projects worldwide use LCC-HVDC systems ranging from 500 to 1500 MW to facilitate the connection of large-scale OWFs in remote host regions to distant load centers [52]. However, an LCC operates only as a rectifier or an inverter, and thus controls either the DC voltage or the DC current [53]. In contrast, a VSC can operate as both a rectifier and an inverter. In other words, a VSC can change the direction of DC current and independently control both active and reactive power. A VSC improves reactive power compensation in the absence of additional compensators, rendering the footprint smaller. The power ratings and power conversion efficiencies of VSCs have continued to increase, particularly when VSCs are implemented as modular multi-level converters (MMCs), although the power ratings are lower, and the power losses are higher than those of an LCC [54]. Recently, a hybrid technology using an LCC and a VSC as interfaces with two remote grids was developed to combine the advantages of the two converter types. A hybrid HVDC system can be further developed as a multi-terminal DC (MTDC) system [55]; for example, LCCs, grid-side VSCs (GSVSCs), and OWF-side VSCs (WFVSCs) are linked via a DC network to connect regional AC grids with OWFs, as shown in Figure 1. This enhances the flexibility of inter-grid power delivery and sharing, facilitating the integration of large-scale OWFs with regional load centers.

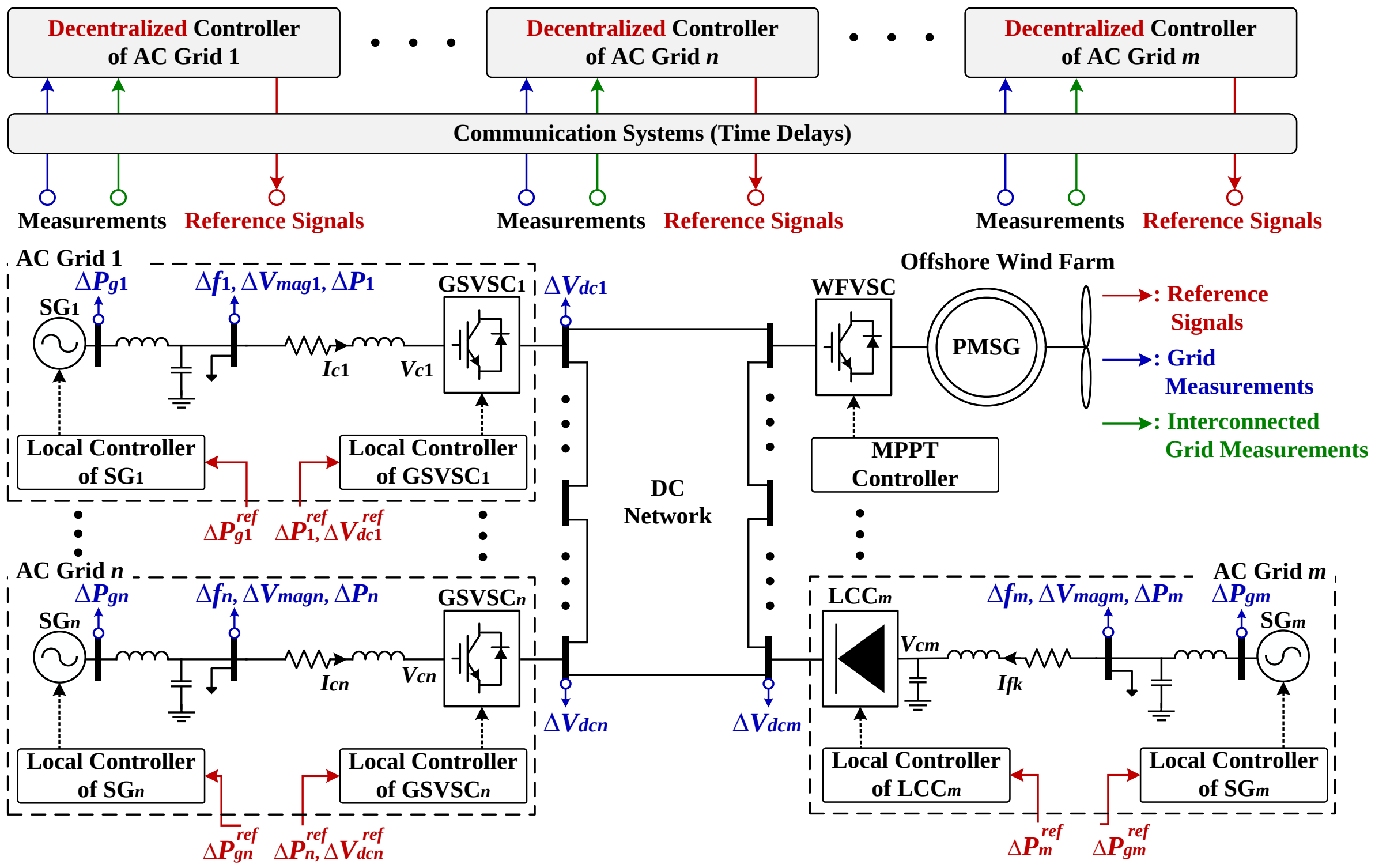

**Figure 1.** A schematic showing optimal control of hybrid MTDC-linked grids and OWFs.

### 3.1.1. Frequency Stability

An HVDC connection causes the frequency in each grid to be affected by any supply-and-demand imbalance, not only in the corresponding grid but also in other grids. Similarly, the intermittent power generation of OWFs triggers frequency deviations in all HVDC-linked grids. A strategy is required for real-time supply-and-demand balancing in HVDC-linked grids; this is essentially the coordinated control of regional power generation and AC-to-DC power transmission in either a centralized or distributed manner. For example, considering inter-grid coupling, Figure 2a shows an optimal control strategy for an LCC-HVDC system in South Korea (i.e., the Jeju–Haenam system [56]). It has a rated power of 150 MW and a rated voltage of 184 kV. The HVDC system delivers surplus power generated at the rectifier terminal of the Haenam grid to the inverter terminal of the Jeju grid via a 100 km DC cable. Each converter terminal includes a converter transformer, 12 thyristor valves, and internal controllers that regulate the firing or extinction angles. Optimal coordinated control of the DC-link voltage and current is achieved in real time to improve frequency regulation (FR) in both the rectifier- and inverter-side networks. Specifically, Figure 3 shows a small-signal model of HVDC-linked grids, wherein the input variables include the DC voltage and current references. The HVDC converters support real-time FR via primary and secondary frequency control (PFC and SFC). For PFC, the converters are integrated with feedback loops to afford frequency- and DC-voltage-power droop control, and also inertial response emulation. A linear quadratic Gaussian (LQG) controller that combines a linear quadratic regulator (LQR) with a Kalman filter is also incorporated into the feedback loops to achieve optimal SFC by minimizing the weighted sum of the instantaneous and accumulated deviations in the DC-link voltage and the frequencies of the rectifier- and inverter-side grids. Figure 4 shows the results of comparative case studies of the proposed and conventional strategies, as described in Table 3. Table 4 shows the corresponding numerical results. The proposed strategy improved real-time FR, when OWF power generation and load demand varied over time, under various conditions of the model parameters, LQG weighting factors, and the inertial emulation and droop control approaches.

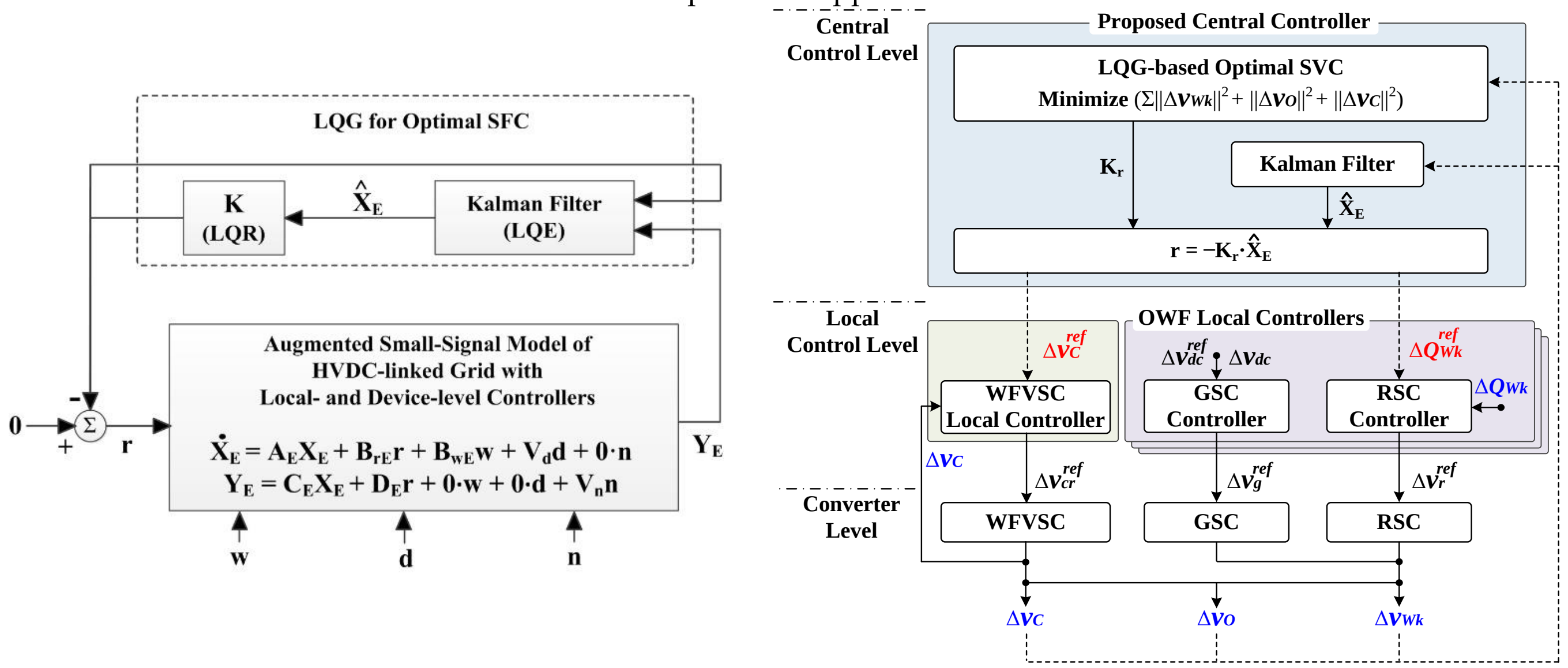

(a) (b)

**Figure 2.** Use of an LQG regulator to generate optimal reference signals for the individual local controllers of the HVDC system and the OWFs: (**a**) frequency control and (**b**) voltage control.

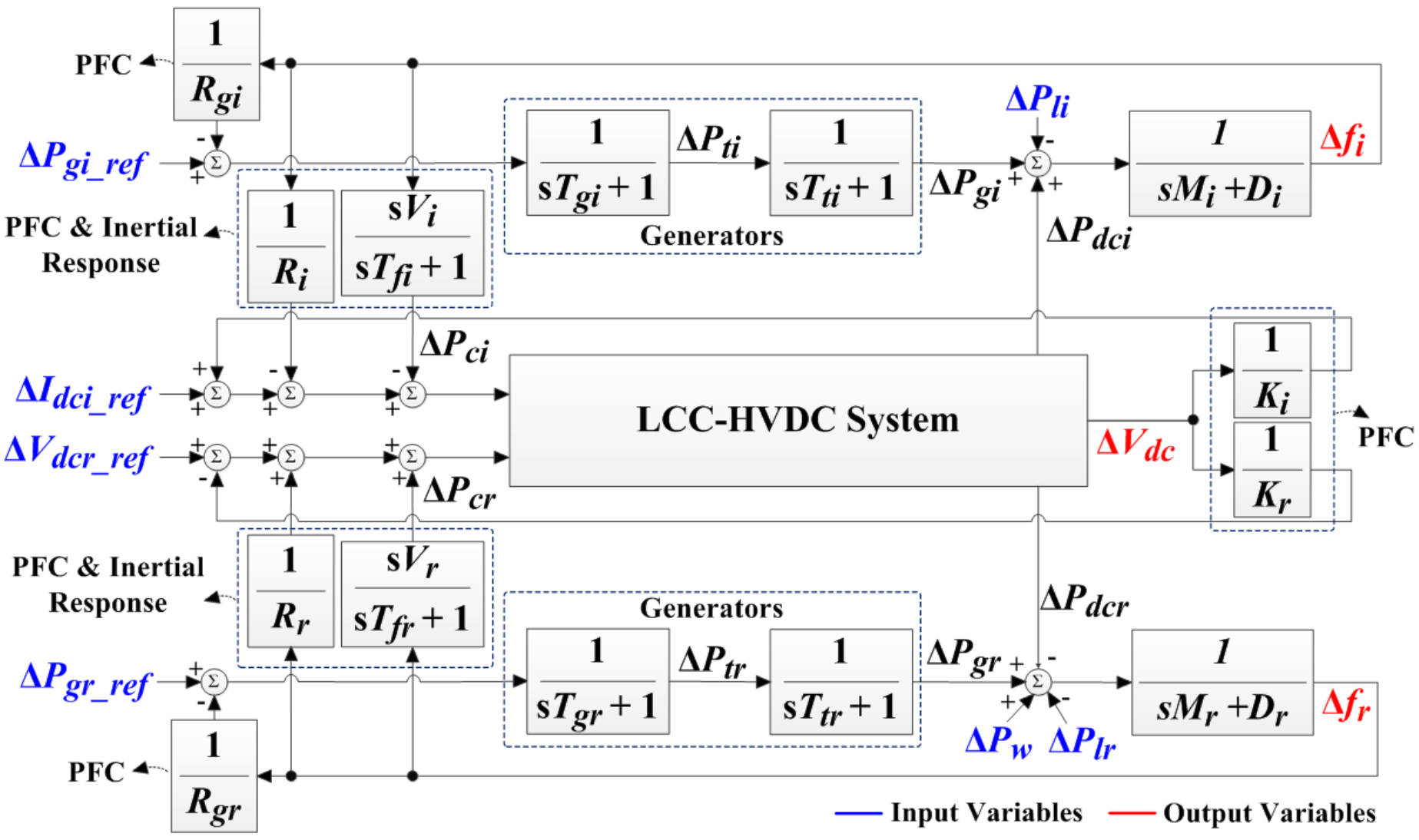

**Figure 3.** A small-signal model of HVDC-linked grids considering the dynamics of the DC link, the converters, and the inner feedback loops.

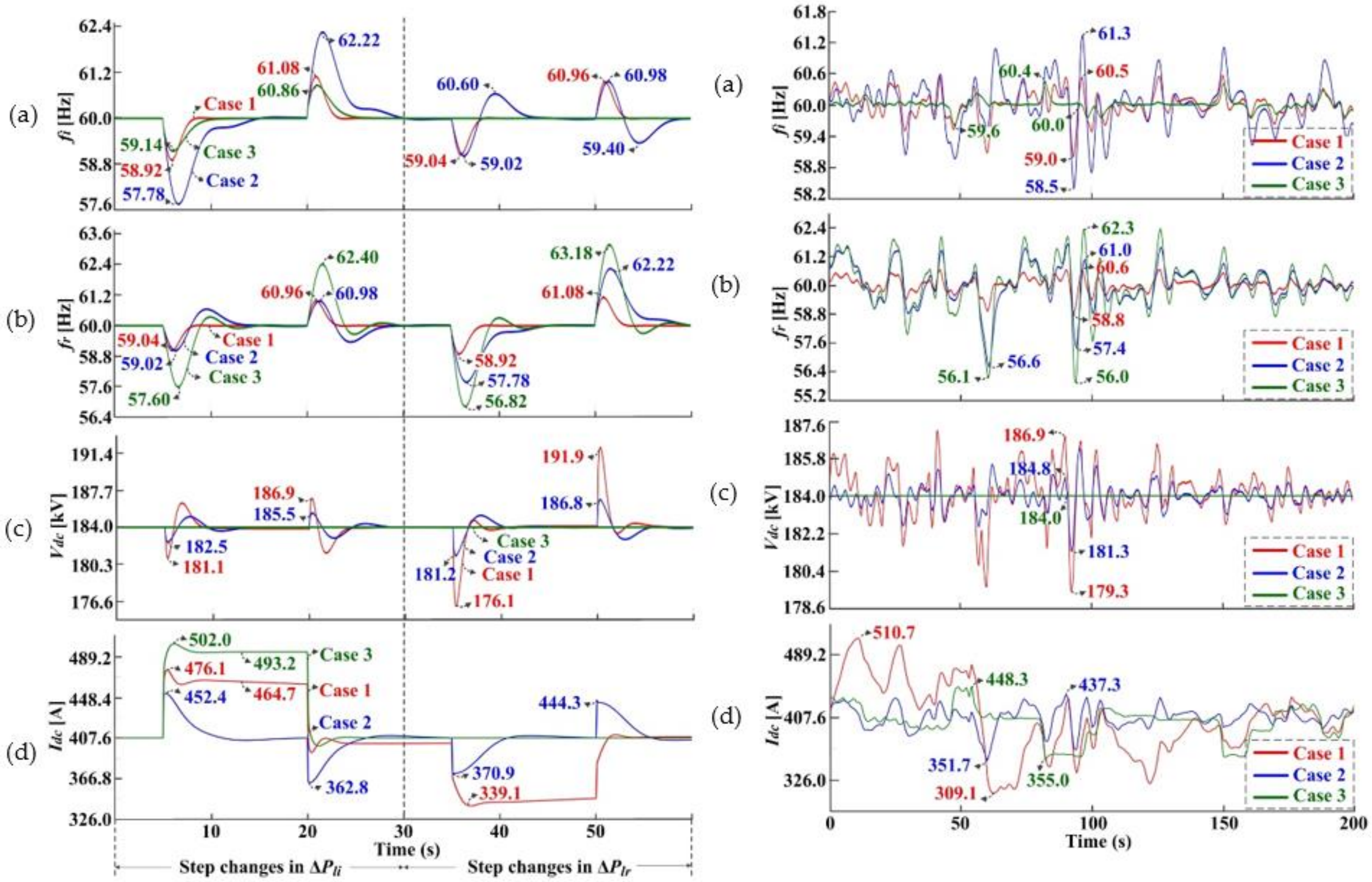

**Figure 4.** Responses to (**left**) step and (**right**) continuous variations in the inverter-side and rectifier-side loads and in the OWF power generation when the LQG strategy (Case 1) and conventional PI-based strategies (Cases 2 and 3) are applied: (**a**) inverter-side frequency $f_i$, (**b**) rectifier-side frequency $f_r$, (**c**) DC-link voltage $V_{dc}$, and (**d**) DC current $I_{dc}$.

**Table 3.** Features of the proposed and conventional strategies.

| HVDC Control Strategies in Figure 4 | | DC-Link Voltage | Control Target Grids | Secondary Frequency Control |
|---|---|---|---|---|
| Proposed | Case 1 | time-varying | both-side grids | LQG |
| Conventional | Case 2 | time-varying | both-side grids | PI |
| | Case 3 | fixed | inverter-side grid | PI |

**Table 4.** Comparisons of the results for the step and continuous response tests.

| Frequency Deviations | | Case 1 | | Case 2 | | Case 3 | |
|---|---|---|---|---|---|---|---|
| | | Individual | Total | Individual | Total | Individual | Total |
| Figure 4 (left) | $\lvert\Delta f_i\rvert_{max}$ [Hz] | 1.08 | 3.16 | 2.22 | 4.44 | 0.86 | 4.04 |
| | $\lvert\Delta f_r\rvert_{max}$ [Hz] | 1.08 | | 2.22 | | 3.18 | |
| Figure 4 (right) | $\Delta f_{i,rms}$ [Hz] | 0.25 | 0.53 | 0.44 | 1.24 | 0.10 | 1.13 |
| | $\Delta f_{r,rms}$ [Hz] | 0.28 | | 0.80 | | 1.03 | |

#### 3.1.2. Voltage Stability

The intermittent nature of wind power also renders it difficult to control offshore grid voltages. In general, OWFs supply power via long feeders with low X/R ratios at medium voltage (MV) levels, leading to considerable deviations in offshore grid voltages as the wind power fluctuates. To address this issue, grid codes have been established that require OWFs to engage in real-time regulation of offshore grid voltages under normal operating conditions. Following the codes, HVDC systems and OWFs can be optimally coordinated to improve real-time voltage stability, as in the case of FR discussed in Section 3.1.1. For example, in [57], the optimal coordination of a VSC-HVDC system and OWFs was achieved using an MPC algorithm to maintain the terminal voltages of the OWFs within an acceptable range. In [58], MPC-based coordination was achieved in a distributed manner. Moreover, Figure 2b shows an optimal real-time strategy for secondary voltage control (SVC) of HVDC-linked OWFs, with consideration of the communication time delays between the central and local controllers [59]. The optimal coordination between the HVDC system and OWFs reduced voltage fluctuations at the PCC, POC, and the terminals of OWFs in the offshore grid. Using a dynamic model of HVDC-linked OWFs, an LQG regulator was designed to generate continuous reference signals ensuring optimal SVC and thus improve robustness against increased communication time delays. Figure 2b also shows the integration of a centralized LQG controller with the local controllers of the HVDC system and OWFs. As depicted in Figure 5, the optimal SVC strategy effectively suppressed voltage deviations in an offshore grid and ensured voltage stability in the offshore grid, even when relatively large time delays occurred. The performance was better than those of conventional no-SVC and PI- and MPC-based SVC strategies, as described in Tables 5 and 6.

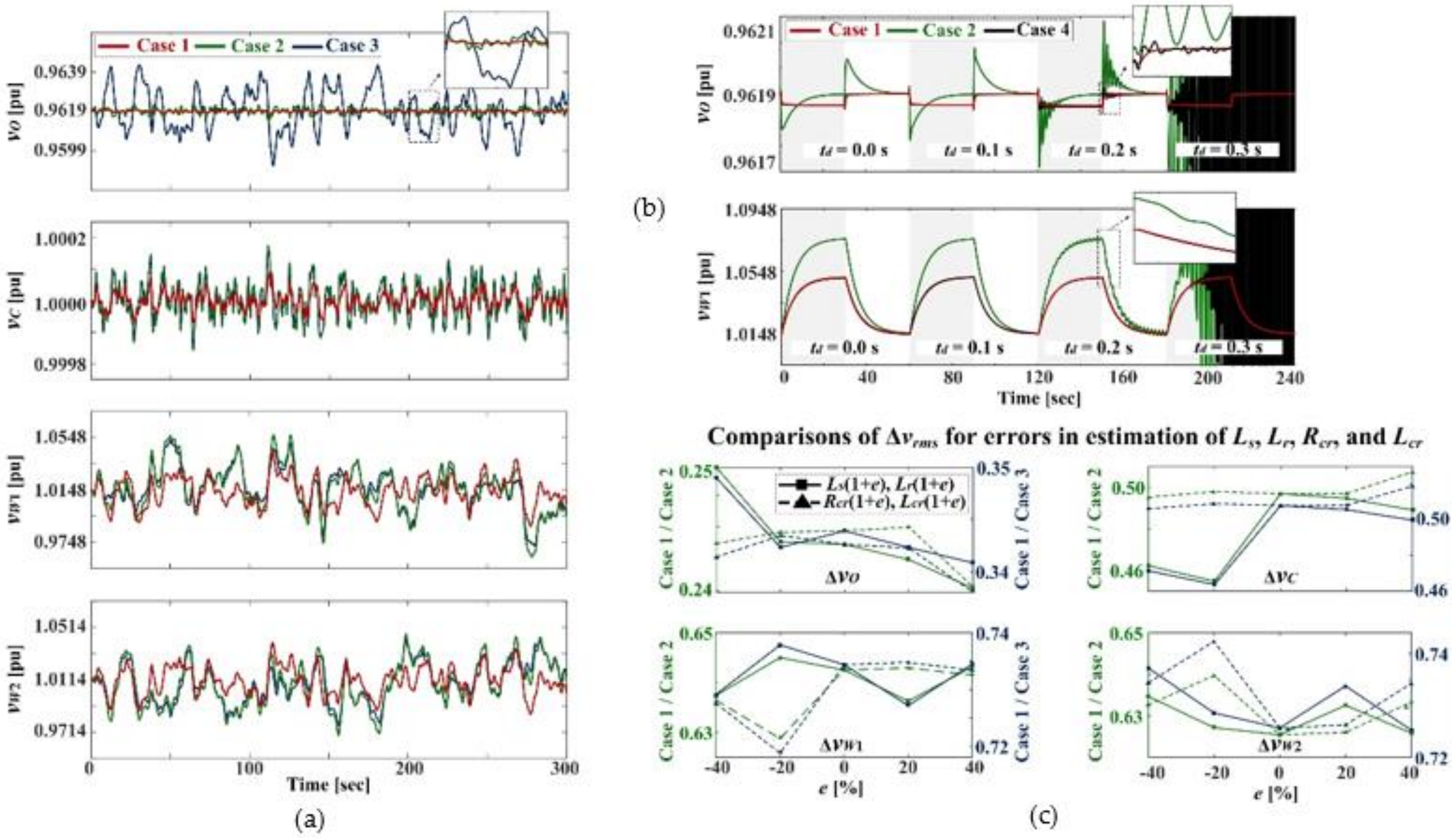

**Figure 5.** Voltage responses to (**a**) continuous variations in OWF power generation, (**b**) a communication time delay $t_d$, and, (**c**) uncertainties in inductance and resistance parameters when a novel SVC strategy (Case 1) and conventional PI- and MPC-based SVC strategies (Cases 2 and 3) were applied. In Figure 5a, $v_O$ and $v_C$ are the POC and PCC voltages, respectively, and $v_{W1}$ and $v_{W2}$ are the OWF terminal voltages.

**Table 5.** Features of the proposed and conventional strategies.

| SVC Strategies | | | Reference Voltages | | |
|---|---|---|---|---|---|
| | | | $v_C^{ref}$ | $v_O^{ref}$ | $v_{Wk}^{ref}$ **(or** $Q_{Wk}^{ref}$**)** |
| Prop. | Case 1 | LQG | time-varying | time-varying | time-varying |
| Conv. | Case 2 | PI | 1 pu | $\Delta v_O^{ref} = 0$ pu | - |
| | Case 3 | No-SVC | 1 pu | - | $\Delta Q_{Wk}^{ref} = 0$ pu, $\forall k$ |
| | Case 4 | MPC | time-varying | time-varying | time-varying |

Gray: centralized control, white: localized control.

**Table 6.** Comparison of continuous response test results.

| Maximum Variations in Figure 5a | $\lvert\Delta v_O\rvert_{max}$ (×10$^{-5}$) | $\lvert\Delta v_C\rvert_{max}$ (×10$^{-5}$) | $\lvert\Delta v_{W1}\rvert_{max}$ | $\lvert\Delta v_{W2}\rvert_{max}$ |
|---|---|---|---|---|
| Case 1 [pu] | 6.4840 | 4.3595 | 0.0363 | 0.0354 |
| Case 2 [pu] | 18.1581 | 5.8330 | 0.0610 | 0.0614 |
| Case 3 [pu] | 348.8569 | 5.8436 | 0.0478 | 0.0474 |

### *3.2. Network Reconfiguration and Distributed Generation*

Climate change affects the technical and economic considerations of the power industry. Extreme weather events, such as floods and storms, increasingly threaten grid reliability [60]. Given the rapid increase in the frequency of such events, network reconfiguration (NR) has received significant attention because it can mitigate adverse effects on grid reliability by changing the network topology using on–off switches (SWs). For example, in Figure 6a, an islanded microgrid (MG) is implemented using the IEEE 37-node test feeder. The MG includes tie SWs (TSWs) and sectionalizing SWs (SSWs) that can adaptively change the MG topology in real time. Initially, the TSWs and SSWs are open and closed, respectively. Figure 6b shows an NR scenario, in which the SW operations trigger variation in the MG topology to restore the critical loads in $LA_1$ and $LA_3$. The non-critical loads in $LA_2$ were shed to ensure sufficient reserve capacity of the distributed generators (DGs) and then recovered after all critical loads became energized.

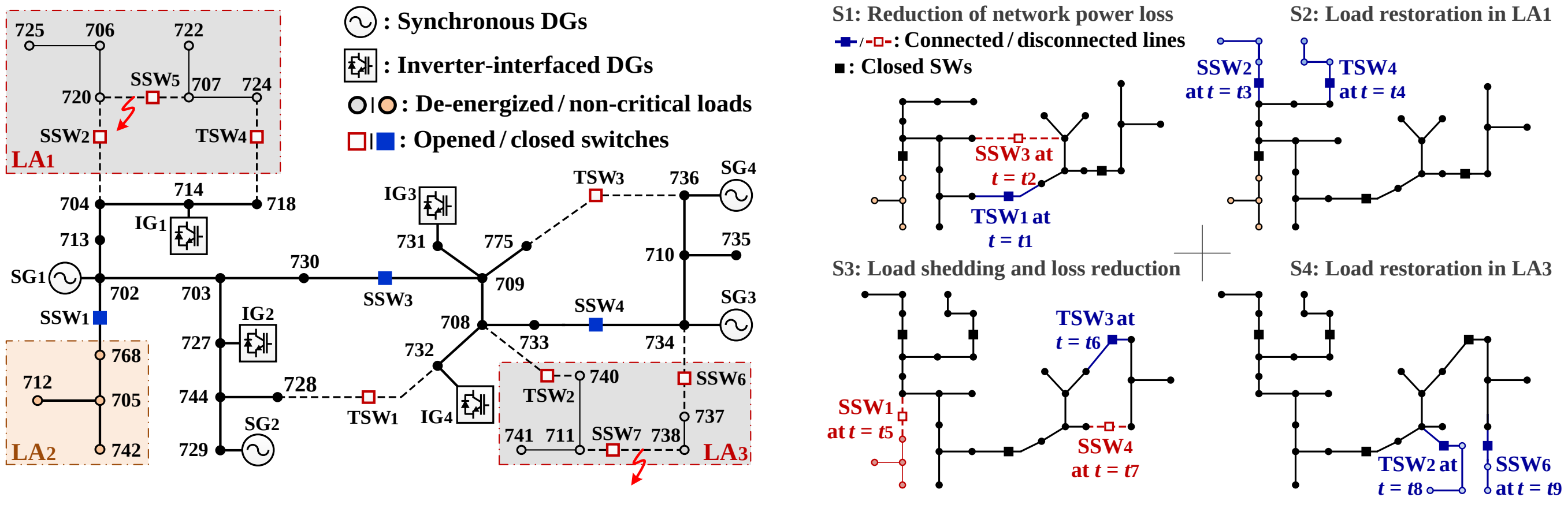

(**a**) (**b**)

**Figure 6.** (**a**) A reconfigurable network with switches and synchronous machine-based and inverter-interfaced generators and (**b**) the variations in the topology that reduce power loss and restore and shed load.

Most NR studies focus on the operational scheduling of SSWs and TSWs in the steady state, for example, to maximize the restored load demand and minimize the restoration time. SSW and TSW scheduling principally considers the steady-state operations of distribution grids and MGs using constant or hourly sampled load demand forecasts before and after NR. During such scheduling, DGs featuring synchronous machines and inverters (referred to as SGs and IGs, respectively) have often been treated as point sources (i.e., PQ or PV nodes) without consideration of the dynamic responses to SW operations. In practice, consecutive SW operations and the resulting abrupt variations in load demand can trigger severe fluctuations in grid frequency and node voltages in the transient state, given the small capacities and low moments of inertia of DGs, which can thus unexpectedly trip DGs and cascade the collapse of network voltages.

Two recent studies [61,62] focused on how real-time control of DGs, working with the SWs, enhanced grid stability and reliability during load restoration. Both studies used DG feedback control loops to ensure that the DG reference power outputs and terminal voltages were determined based on information from the SWs. The feedback loops were activated only after the load demand variation caused by NR significantly affected the MG operation. These studies motivated the development of new control strategies; DGs can pre-emptively compensate for the load variations given that, in general, NR initiation is controlled. The pre-emptive control of power grids has become increasingly possible as modeling and parameter estimation techniques have evolved over time.

### 3.2.1. Dynamic NR Modeling and Analysis

When seeking to establish pre-emptive DG control, the research gap in NR modeling accuracy must first be considered. NR has usually been modeled employing the load to be restored or shed [61], rather than as a change in the network topology per se. This can compromise the estimation of the dynamic responses of DGs and loads to the SW operations of NR-aided load restoration, preventing pre-emptive DG control in a reconfigurable network. A frequency response rate (FRR) model has often been used to estimate frequency dips caused by sudden load pickups. However, the transient variations in bus voltages and line losses attributable to NR cannot be analyzed using FRR models. Figure 7 compares a new NR modeling method [63] to a conventional method [61]. To improve estimation accuracy, the new method incorporates SW operations directly, and the corresponding load variations indirectly, into changes in the node currents and voltages of a reconfigurable network. In the new method, step variations in the *dq*-axis node injection currents attributable to SW operations are identified and analyzed with respect to their effects on the dynamic responses of the low-voltage network. Using a defined test bed (Figure 6), Figure 8 shows the results of a case study, exploring the effectiveness of the new method in terms of improving the estimated accuracies of MG frequency and voltage variations attributable to NR. The new method has a great potential, for example, for application to the optimal MPC of DGs, which is difficult to achieve via either conventional modeling method or a trial-and-error approach with numerical simulation. The test grid in Figure 6 includes not only conventional synchronous machine-based DGs, but also inverter-interfaced VRE sources, and confirms the applicability of the NR modeling/analysis method to future VRE-dominated grids.

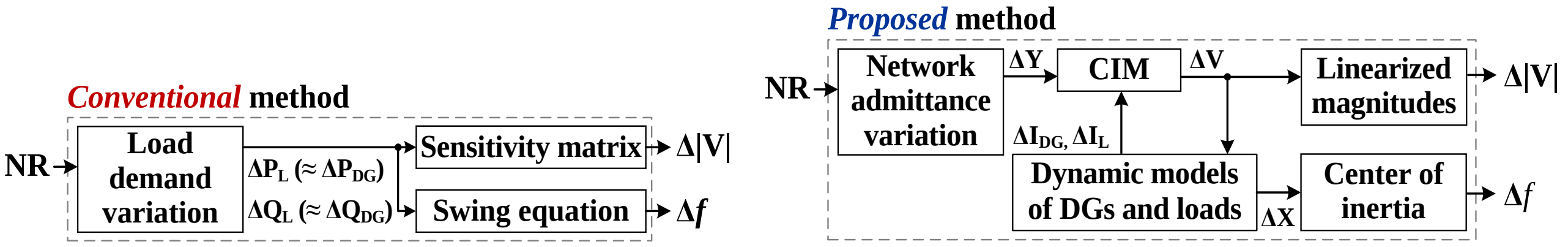

(**a**) (**b**)

**Figure 7.** The (**a**) conventional and (**b**) new methods for NR modeling.

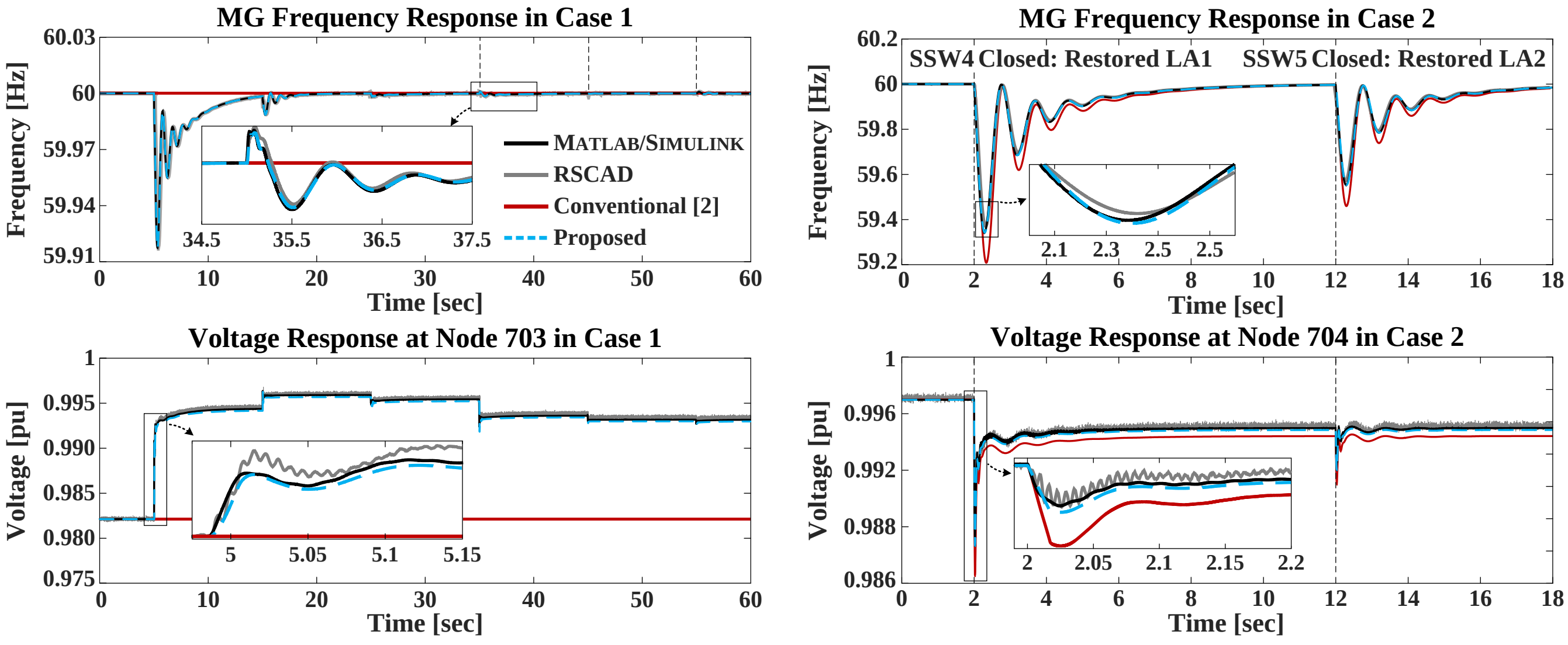

(**a**) (**b**)

**Figure 8.** The grid frequency and voltage profiles of the new, conventional, and comprehensive models of NR: (**a**) Case 1 (without NR-aided load shedding/restoration) and (**b**) Case 2 (with NR-aided load restoration).

### 3.2.2. Real-Time DG Control in Coordination with Scheduled SW Operations

Existing DG control relies on feedback loops. Uncertainties in the estimates of DG and load modeling parameters should be considered. Otherwise, the DG-mediated control of grid frequency and voltages can be ineffective; stable and reliable grid operation is compromised. Figure 9 shows a new strategy for real-time FR of a reconfigurable MG; the feedforward control of SG and IG power outputs is coordinated with SW on–off operations to reduce frequency deviations attributable to NR-aided load restoration [64]. The feedforward frequency controllers (FFCs) were developed with the aid of an analytical dynamic model of the reconfigurable MG, as discussed in Section 3.2.1. The FFCs were then integrated with the PFC and SFC feedback loops (Figure 10) to ensure power sharing among the DGs and a supply-and-demand balance in the steady state, respectively. Figure 11 shows the results of small-signal analytical and simulation case studies. The proposed control strategy was more effective and robust than the conventional feedback-only strategies; please see Tables 7 and 8 for the features of the proposed and conventional strategies and the comparisons of their numerical results, respectively.

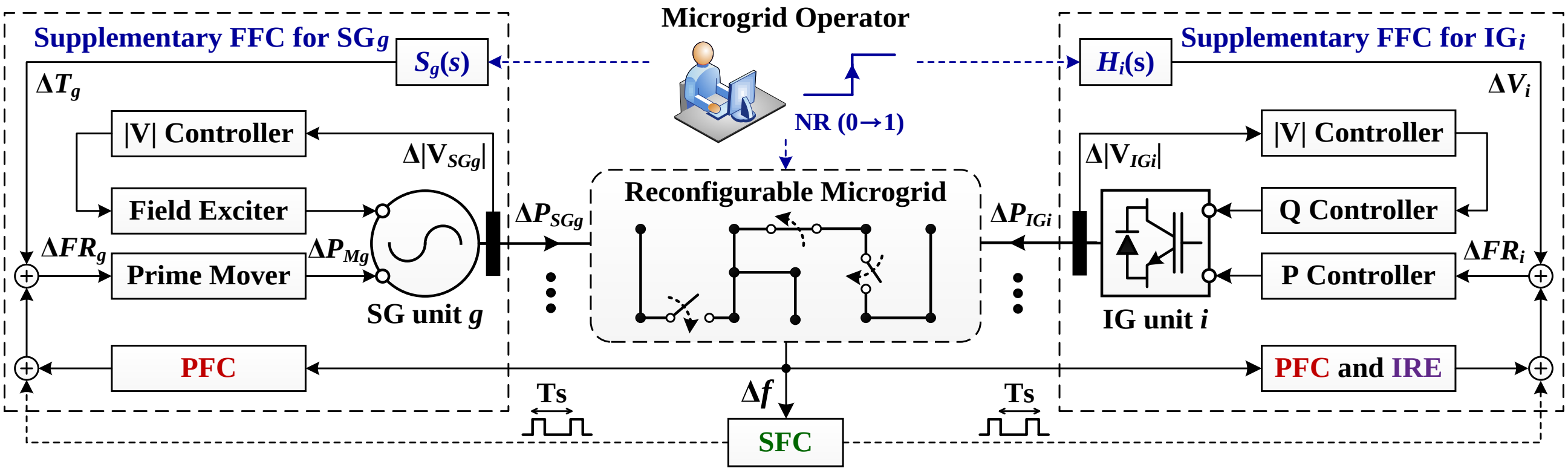

**Figure 9.** Schematic of a proposed FR strategy for an islanded, reconfigurable MG.

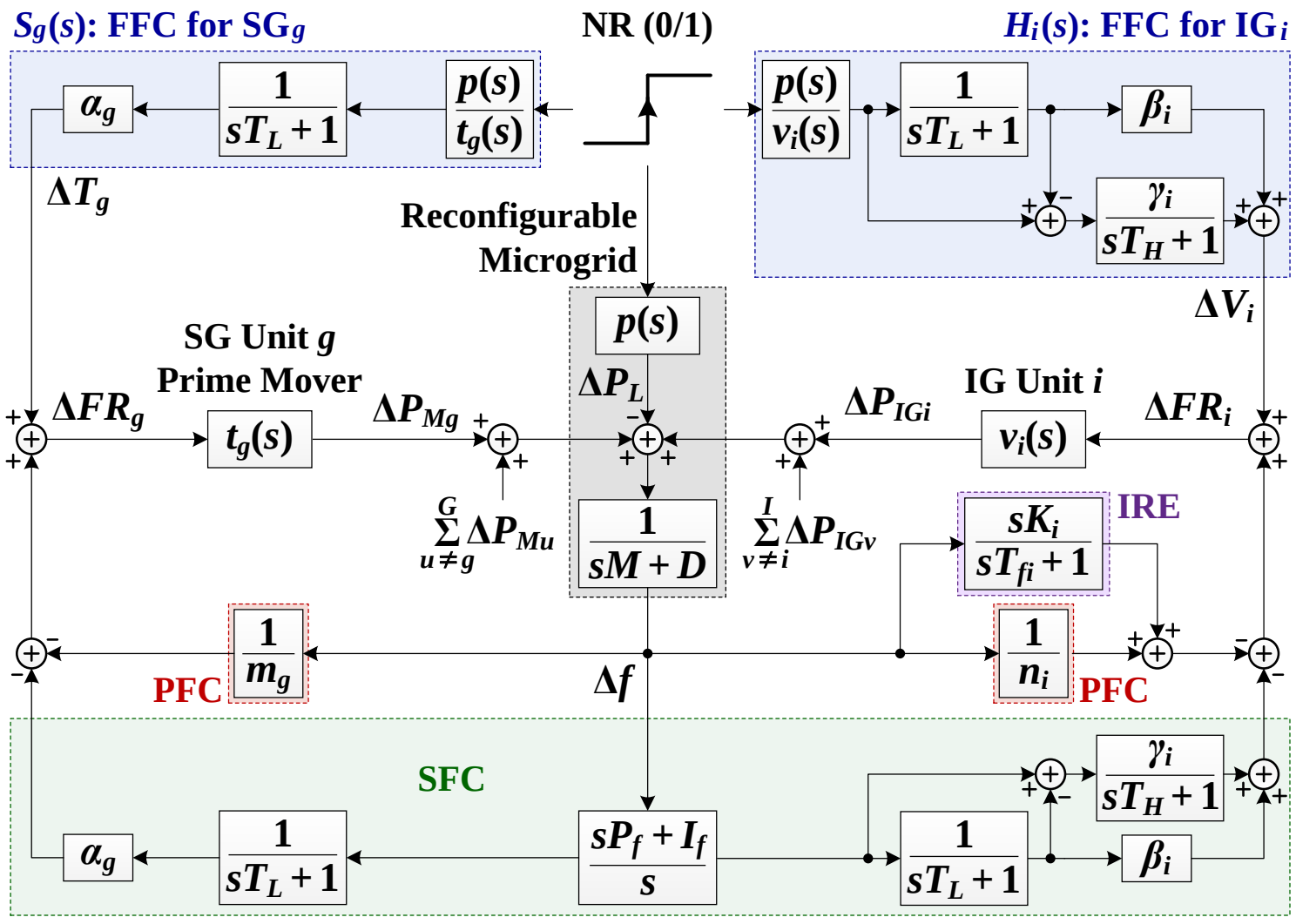

**Figure 10.** A small-signal model of an islanded reconfigurable MG with supplementary FFCs showing the feedback loops for the inertia response emulation and primary and secondary frequency control.

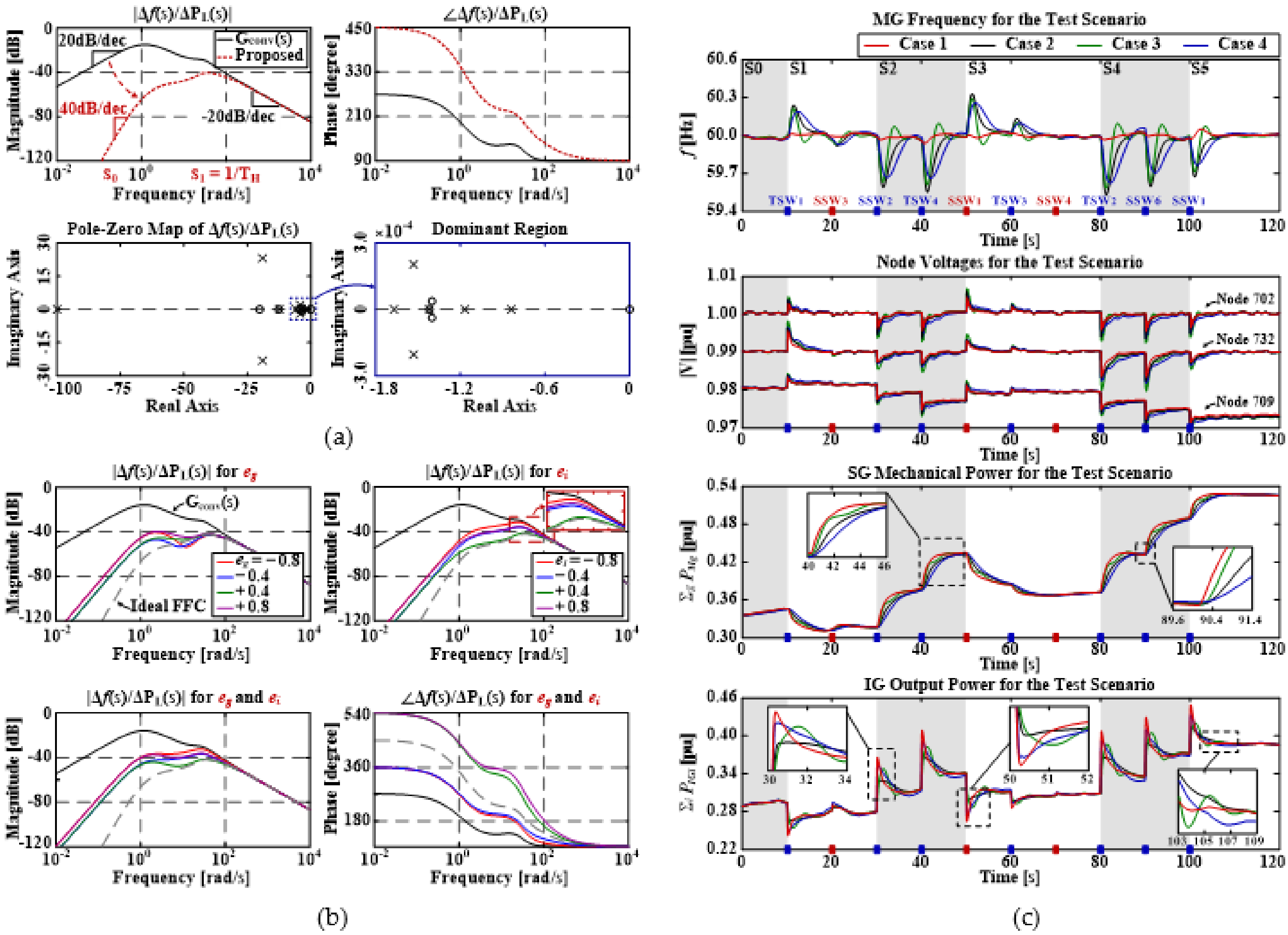

**Figure 11.** (**a**) Bode plots of the $\Delta f(s)/\Delta P_L(s)$ values for the new FR strategy (Case 1) and conventional FR strategies (Cases 2–4); (**b**) Bode plots of $\Delta f(s)/\Delta P_L(s)$ values with the errors in DG parameter estimates; and (**c**) comparisons of $f$, $|\mathbf{V}|$, $\Sigma_g P_{Mg}$, and $\Sigma_i P_{IGi}$ (from top to bottom) during NR-aided load restoration.

**Table 7.** Features of the proposed and conventional FR strategies.

| FR Strategies | | SFC, PFC, and IRE Gains |
|---|---|---|
| Proposed | Case 1 | Set as default values in [64] |
| Conventional | Case 2 | Set as default values in [64] |
| | Case 3 | Increasing SFC gains ($P_f = 3$ and $I_f = 6$) |
| | Case 4 | Increasing PFC and inertia gains ($m = 0.30$, $n = 0.05$, and $K = 15$) |

**Table 8.** Comparisons for the continuous load variations.

| Comparison Factors in Figure 11c | | Proposed (Case 1) | Conventional | | |
|---|---|---|---|---|---|
| | | | Case 2 | Case 3 | Case 4 |
| $\Delta f_{pk}$ | [Hz] | 0.134 | 0.825 | 0.716 | 0.613 |
| $\Delta f_{rms}$ | [Hz] | 0.026 | 0.183 | 0.159 | 0.163 |
| $\Delta P_{M,rms}$ | [pu] | 0.163 | 0.130 | 0.147 | 0.118 |
| $\Delta P_{IG,rms}$ | [pu] | 0.160 | 0.126 | 0.145 | 0.142 |

Similarly, Figure 12a shows a new strategy for the real-time voltage regulation (VR) of a reconfigurable MG [65]. Optimal feedforward control of the terminal voltages of SGs and IGs is coordinated with SW operations to pre-emptively mitigate transient voltage deviations caused by NR. The dynamic responses of bus voltages to NR are estimated using an analytical model of a reconfigurable network, discussed in Section 3.2.1. The responses are integrated to afford robust optimization of the feedforward voltage controllers (FVCs) that minimize terminal voltage deviations. Uncertainties in the estimates of DG modeling parameters and load demands are considered during optimization, further improving the robustness of FVC operation. The FVCs are incorporated in parallel with the existing feedback control loops, as are the FFCs (Figures 9 and 10); steady-state deviations in DG terminal voltages are eliminated. Figure 12b shows the results of a case study, verifying that the new VR strategy (i.e., Cases 1 and 2 in Table 9) was better than the conventional PI- and robust controller-based strategies (i.e., Cases 3 and 4 in Table 9). Table 10 lists the corresponding numerical results.

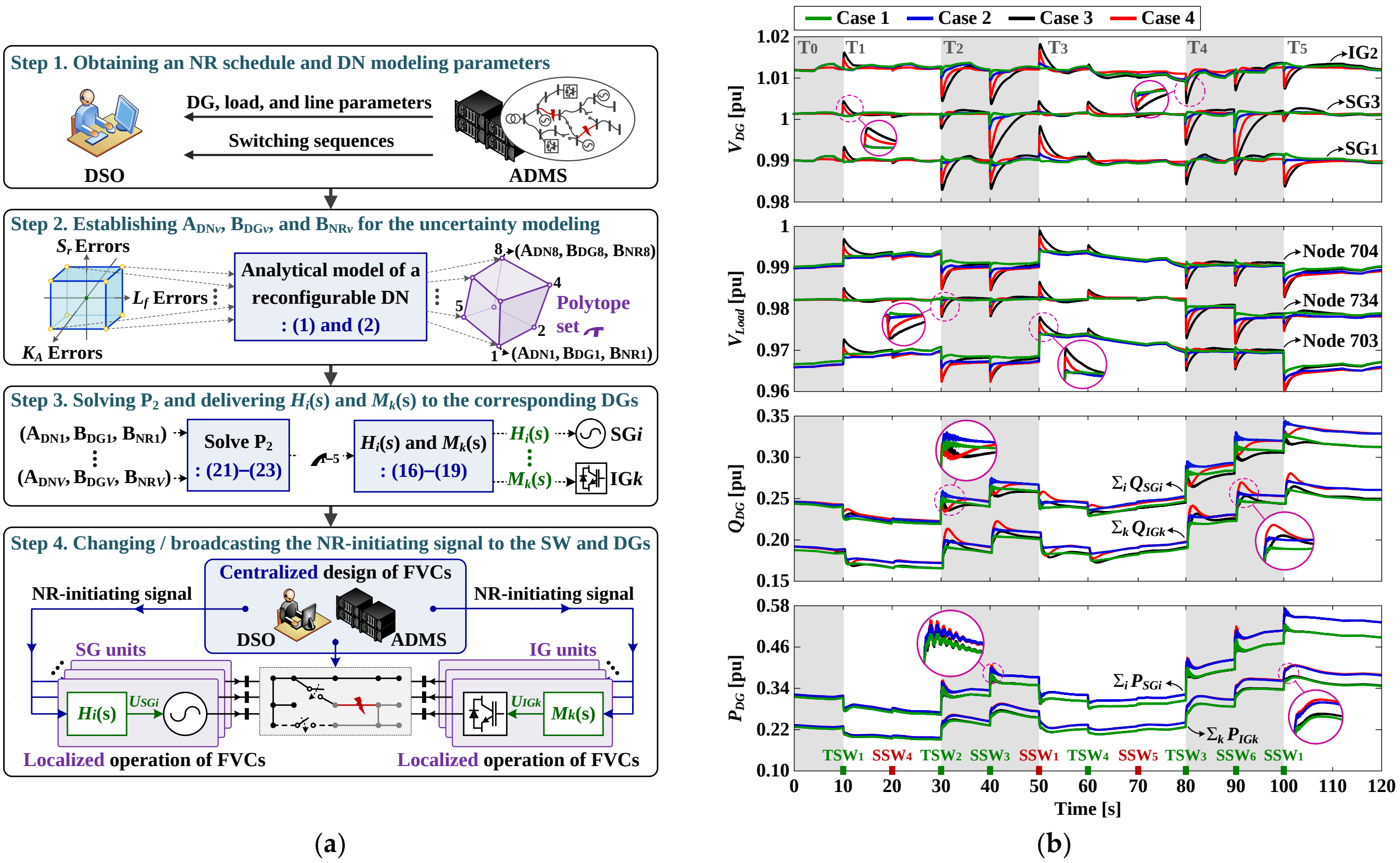

**Figure 12.** (**a**) A flowchart for implementation of the proposed FVCs and (**b**) comparisons of $V_{DG}$, $V_{Load}$, $Q_{DG}$, and $P_{DG}$ (from top to bottom) between the new proposed VR strategy (Cases 1 and 2) and conventional VR strategies (Cases 3 and 4) for NR-aided load restoration.

**Table 9.** Features of the proposed and conventional strategies.

| VR Strategy | | Description |
|---|---|---|
| Proposed | Case 1 | No uncertainties in the parameter estimates |
| | Case 2 | 30% uncertainties in the parameter estimates |
| Conventional | Case 3 | PI-based output feedback loop |
| | Case 4 | Robust state feedback loop |

**Table 10.** Comparisons for the continuous load variations.

| Comparison Factors | | Proposed | | Conventional | |
|---|---|---|---|---|---|
| | | Case 1 | Case 2 | Case 3 | Case 4 |
| $\Delta V_{rms,avg}$ | [×10⁻³ pu] | 1.564 | 1.816 | 6.684 | 3.808 |
| $\Delta V_{pk,max}$ | [×10⁻² pu] | 0.962 | 1.163 | 2.741 | 2.418 |
| $\Sigma_i \Delta Q_{SGi,rms}$ | [pu] | 0.118 | 0.137 | 0.111 | 0.131 |
| $\Sigma_k \Delta Q_{IGk,rms}$ | [pu] | 0.082 | 0.099 | 0.075 | 0.092 |

### *3.3. Balancing Service Provision by Energy Storage*

#### 3.3.1. Real-Time Control and Analysis of Electrical Storage

Apart from HVAC units, behind-the-meter electrical energy storage is a promising DR resource. For example, many residential and commercial buildings are equipped with wall-mounted batteries and EV charging stations. Using such batteries, buildings can provide ancillary services aiding supply-and-demand balancing in real time, thus mitigating the balancing reserve requirements imposed on dispatchable generator units. Electric batteries exhibit high energy densities and rapid dynamic responses, rendering them suitable for high-power cyclic operations. Battery life reduction is caused by deep discharge during driving, not by small swings in the SOC caused by direct load control (DLC) [66].

The distribution of the DLC signal to individual EV batteries requires careful consideration; grid operators will encounter difficulties when simultaneously controlling a large number of EV batteries. A new type of building electrical system, and building- and EV-level controllers, are needed to achieve efficient distribution of DLC signals and hence reliable provision of ancillary services. In [48], an electrical system for a commercial building (Figure 13) coordinated the building-level controllers and the EV charging controllers. Thus, the building served as an intermediate aggregator between the grid operator and multiple EV owners and, consequently, operated as a large-scale stationary battery. This facilitated centralized and decentralized control of the EV batteries. Figure 14 shows a new FR strategy using a building's electrical system, wherein the EV batteries compensate for rapid load variations via DLC, reducing the frequency deviation and the required reserve capacity of synchronous generators [67]. In practice, SFC signals contain long-term variations in DC offset, motivating the design of high-order filters to remove that offset. The proposed FR strategy was implemented using a laboratory-scale microgrid with a real DLC-enabled battery pack, as shown in Figure 13b. The laboratory-scale microgrid corresponds to the IEEE 34-node test feeder with the total capacity of 25 MVA; the detailed specifications of the test feeder were provided in [67]. Figure 15 shows the results of experimental case studies; the FR strategy using EV batteries maintained the real-time supply-and-demand balance more effectively than did a conventional strategy (with no battery) as the feedback controllers, battery capacities, and the maximum PV power generation varied.

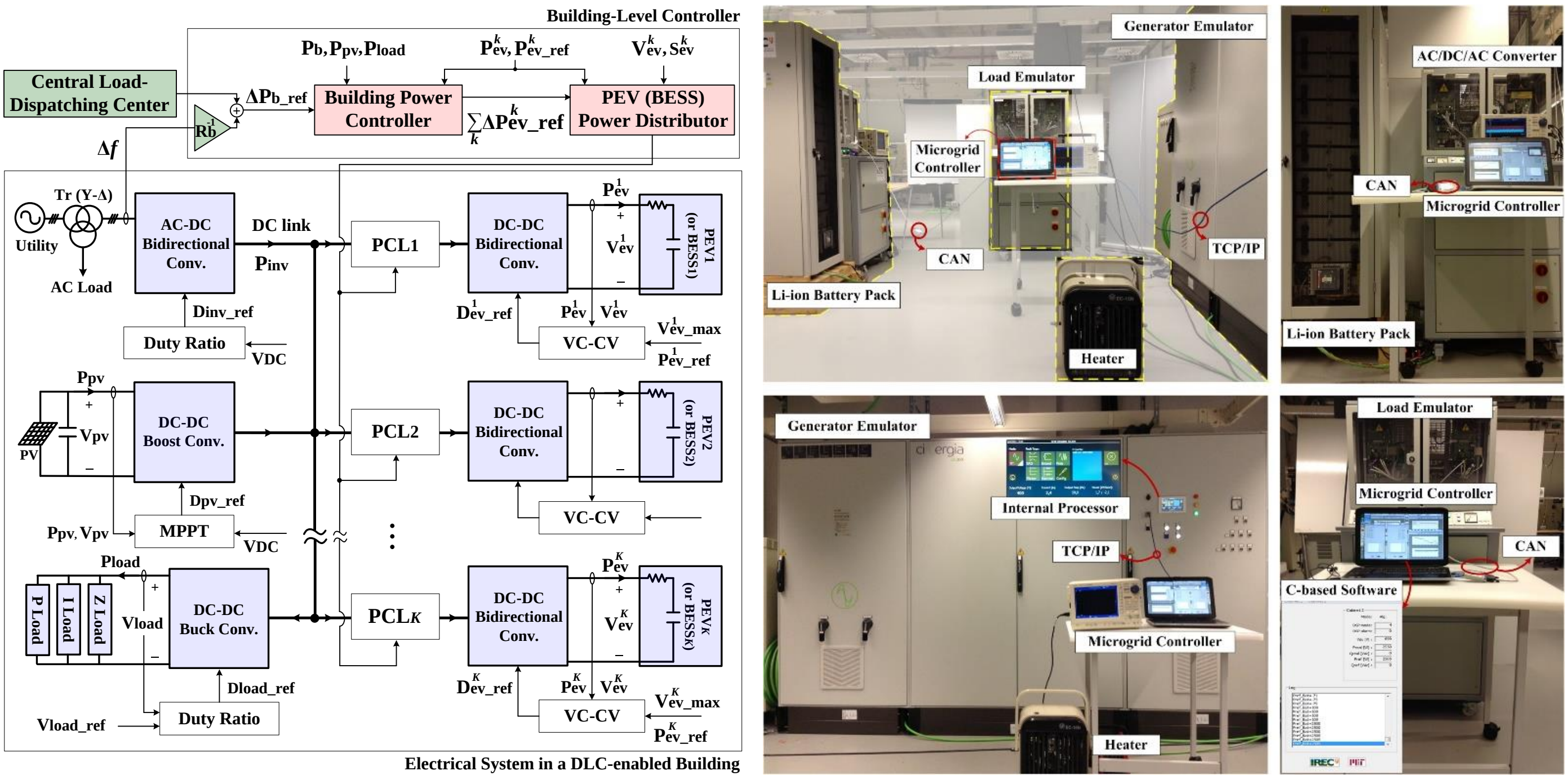

(**a**) (**b**)

**Figure 13.** (**a**) A schematic of a new electrical system inside a commercial building and (**b**) the experimental setup of an MG including a generator emulator, a battery pack, and a load emulator.

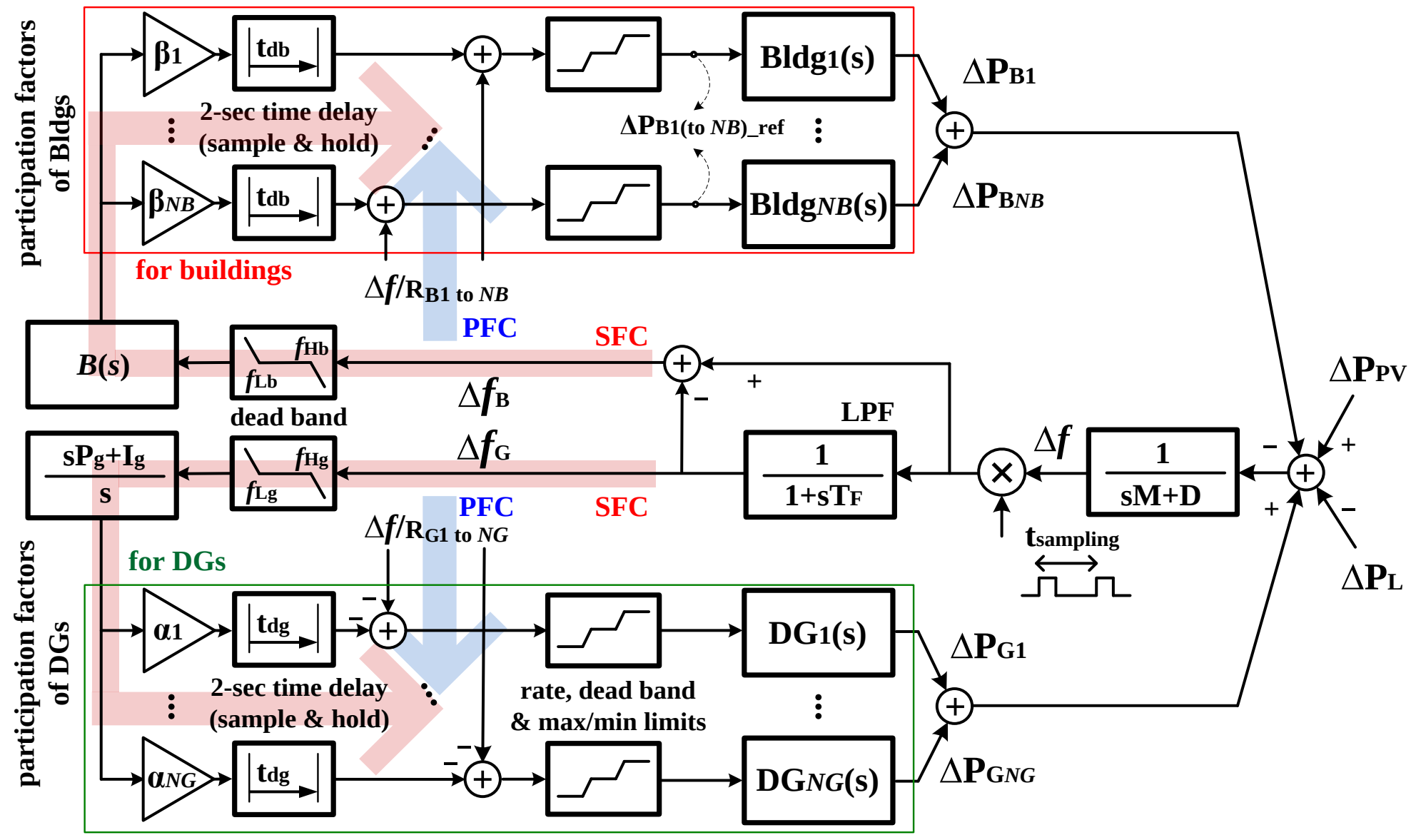

**Figure 14.** A block diagram of the ancillary service provision by buildings including EV batteries.

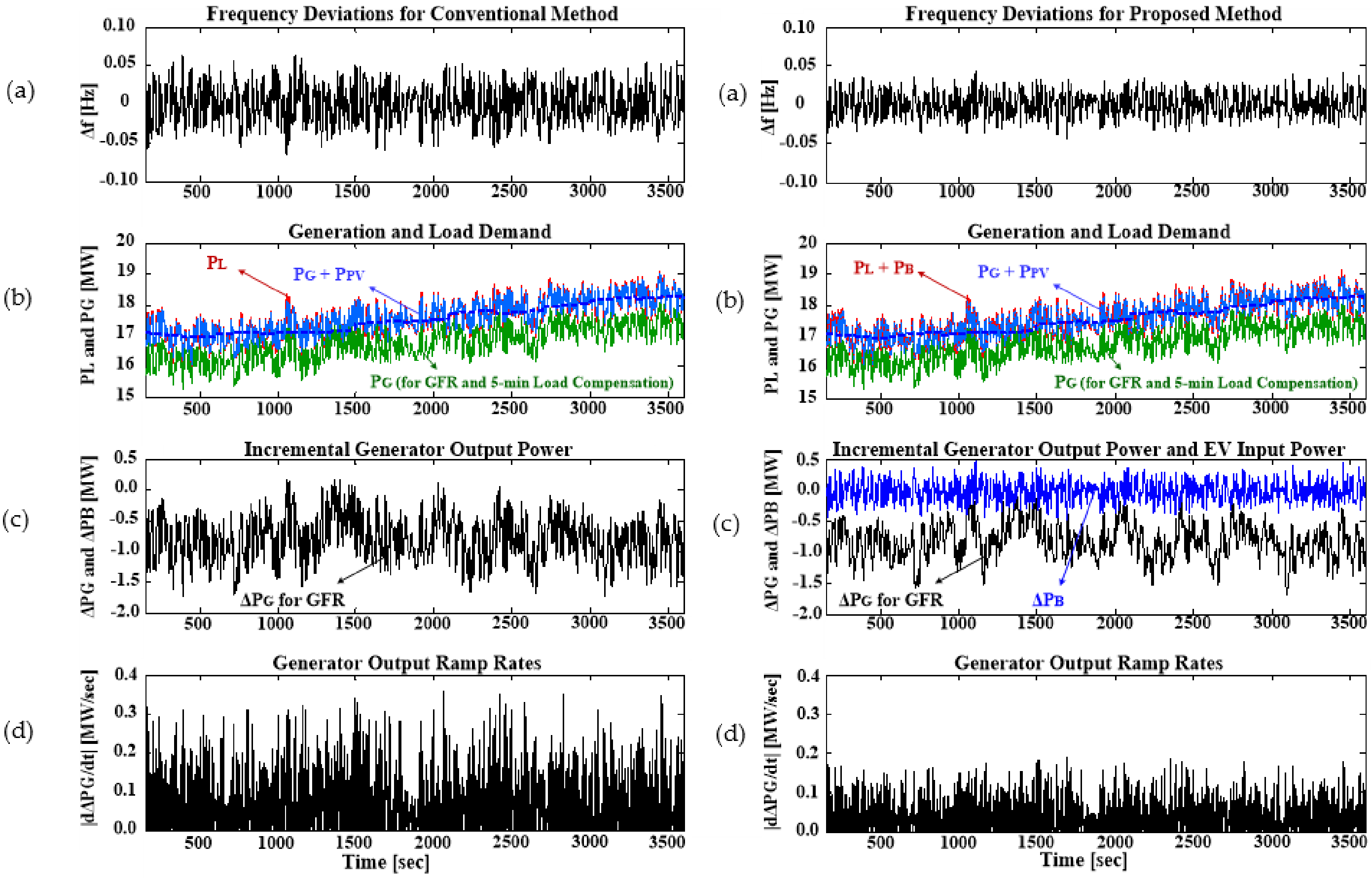

**Figure 15.** Experimental case study results for the (**left**) conventional and (**right**) proposed FR strategies: (**a**) frequency deviations $\Delta f$, (**b**) load demands $P_L$, dispatchable generation $P_G$, and total generation $P_G + P_{PV}$, (**c**) variations in generator output power $\Delta P_G$ and in battery input power $\Delta P_B$, and (**d**) dynamic ramping rate of generation $|d\Delta P_G/dt|$.

### 3.4. Decentralized State Estimation

Network state estimation (SE) processes noisy or inaccurate data and evaluates the true operating state, and hence, the stability, of a grid. SE is essential, for example, in terms of optimal power flow, contingency analysis, and corrective control. Many studies have sought to decentralize SE and thus, develop hierarchical SE (HSE) and distributed SE (DSE). During HSE, each local estimator acquires SE data only for a sub-area. A central estimator calculates global estimates by adjusting all local SE results. In contrast, DSE avoids the need for central coordination. In the literature, DSE was explored using the alternative direction method of multipliers (ADMM), by which an agent acquires and shares local state estimates only with neighboring agents. As grid communications continue to improve, local estimates can readily be shared within and between sub-areas. The ADMM can be achieved using only global observability [68] (rather than local observability), thus reducing the number of measurements required.

#### 3.4.1. Hybrid Decentralized SE Using PMUs

Phasor measurement units (PMUs) have significant advanced SE. Hybrid DSE blends the measurements of supervisory control and data acquisition (SCADA) systems and PMUs via single- and multi-stage approaches. In the single-stage approach, a local estimator integrates the SCADA and PMU data into a single hybrid measurement function. A weighted least-squares (WLS) optimization problem is then formulated that functions and is solved in a distributed manner. The sampling rates of SCADA and PMU measurements differ and thus, the single-stage approach often requires substantial modification of existing SCADA-based SE systems [69]. Recently, the multi-stage approach was considered a good alternative. The SCADA and PMU data are subjected to separate measurement functions in a hybrid DSE. This can proceed in either a sequential or parallel manner. Most recent studies (e.g., [70–72]) have focused on the sequential approach; PMU estimators first obtain and deliver SE results to SCADA estimators, or vice versa. For example, in [70,71], the PMU estimators first obtained the SE models and then provided them to the SCADA estimators, along with information on the boundary states and measurement couplings. The SCADA estimators then used linear SE models to reduce the sizes and complexities of non-linear SE problems. However, the sequential approach is less time efficient than the parallel approach [72], in which SCADA- and PMU-based systems acquire state estimates separately and calculate their weighted sum to determine the final SE results.

#### 3.4.2. Incorporation of Decentralized Bad Data Processing into Decentralized SE

The increasing support of information and communication technology (ICT) enables the conversion of legacy power grids into smart grids, wherein the tasks of grid operation, control, and estimation are achieved using the continuous flows of data measured from remote power facilities. However, ironically, the development of ICT increases the surfaces that malicious actors can attack. In particular, a power grid spans a wide geographic area, and hence the public and private communications between remote facilities and control centers are likely to increase the numbers of sites that hackers and attackers can readily access to pollute measured data with missing data and bad data (BD). This will cause disruptions to the operational stability of power grid and further large-scale power outages, leading to serious and harmful consequences in society and economy. It was reported in [73,74] that various types of cyberattacks successfully occurred in essential parts of the power grid. This implies that cyber security should be recognized as a critical issue, as comprehensively discussed in [75,76], when moving toward power grid decarbonization.

BD acquired from critical measurements reduce the accuracy and computational efficiency of BD processing (BDP) methods, for example, using largest normalized residual tests (LNRTs), least absolute values, and sparse $l_1$-relaxation. Multiple BD conforming are likely to further decrease the accuracy and efficiency [77]. This is because the advantages afforded by PMU measurements have not been fully exploited, compromising the real-world hybrid DSE performance. In [77,78], a phasor-aided SE (PHASE) method was used to improve BDP performance via more intensive cross-validation of the SCADA and PMU data. However, the PHASE method was implemented only in a centralized manner.

Figure 16 compares a common, centralized hybrid SE strategy to a new hybrid DSE strategy [79]. In the latter, the PHASE method is integrated with ADMM-based DSE to improve BDP performance and thus also the DSE. Here, we term the new hybrid DSE, "decentralized PHASE (DPHASE)"; this was developed using a parallel multi-stage approach. Specifically, in each sub-area, the SCADA- and PMU-based estimators obtain the local SE results in parallel using SCADA and PMU measurements, respectively, and the corresponding measurement functions. The local estimators need exchange only small numbers of state estimates with their neighboring estimators, exploiting the ADMM. Given the network topology and measurement locations, the estimators then extend the sets of SCADA- and PMU-based state estimates in a manner that includes the same entries. The final estimates are determined based on fusion of the extended sets. Figure 17 and Tables 11 and 12 show the absolute error and average maximum absolute error (AMAE) results of comparative case studies when conducted on IEEE 14-, 118-, and 1062-bus networks. All the test networks include sub-areas that are locally unobservable to the PMU-based estimators. Specifically, the 14-bus network is divided into four sub-areas [79]. The SCADA systems acquired the voltage magnitudes, power injections, and line power flows at measurement points 4, 5, and 16, respectively. The PMUs measured both the magnitudes and phase angles of the voltages and incident currents at buses 2, 6, 7, and 9. The locations were optimally determined to achieve global observability using the minimum number of PMUs. Similarly, the IEEE 118-bus network was divided into five sub-areas, as shown in Table 13. The PMUs and SCADA sensors were located optimally and arbitrarily, respectively, such that the network was observable globally. The IEEE 1062-bus network includes nine sub-areas, each of which corresponds to the IEEE 118-bus network. The locations of the PMUs and SCADA sensors in each sub-area are the same as those in the 118-bus network.

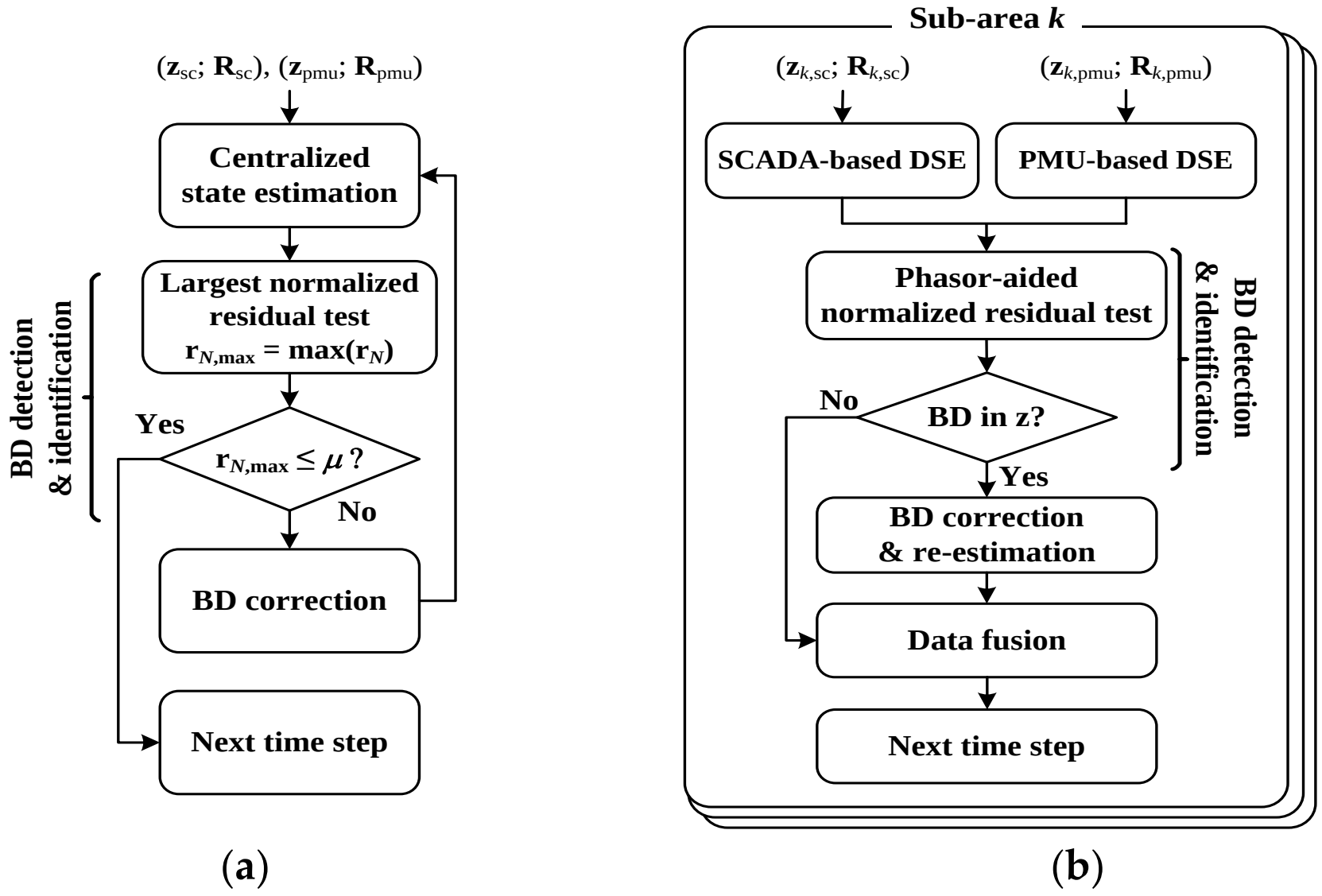

**Figure 16.** Flowcharts of (**a**) a conventional, centralized hybrid SE strategy and (**b**) a proposed, decentralized PHASE strategy, both of which include BD detection, identification, and correction steps.

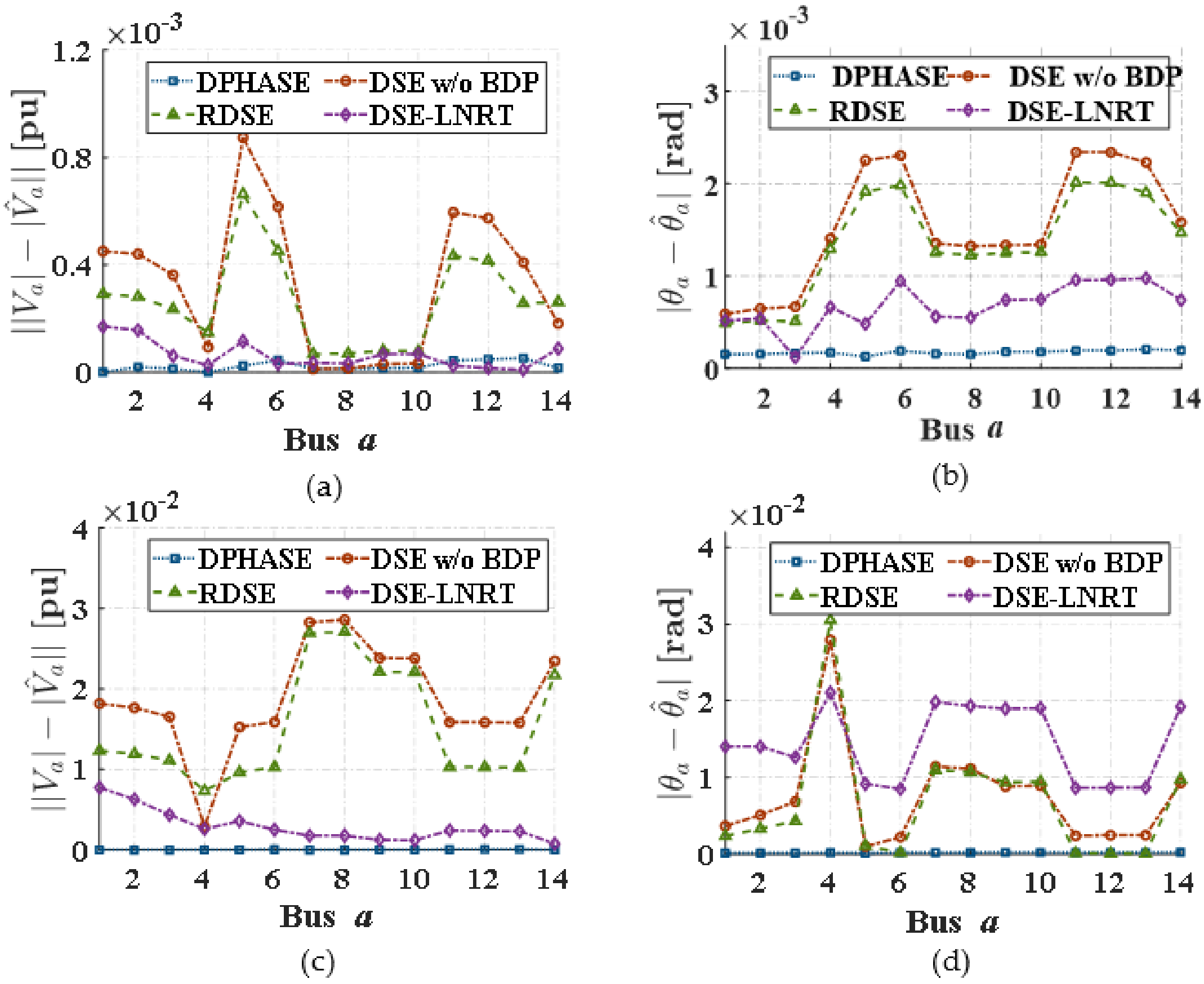

**Figure 17.** Comparison of absolute SE errors: (**a**,**b**) voltage magnitudes and phase angles for the BD in the SCADA measurements and (**c**,**d**) voltage magnitudes and phase angles for the BD in the PMU measurements.

Compared to conventional DSE strategies, the DPHASE strategy improves the accuracy and robustness of SE results under various conditions (the grid size; and the location, magnitude, and numbers of bad data were varied). Moreover, Figure 18 compares the convergences to the optimal SE results of the proposed DPHASE strategy and the conventional DSE-LNRT strategy, showing that DPHASE led to a smaller difference between the actual and estimated states for all areas and buses when the ADMM convergence threshold $\varepsilon_{ADMM}$ was set to be sufficiently small.

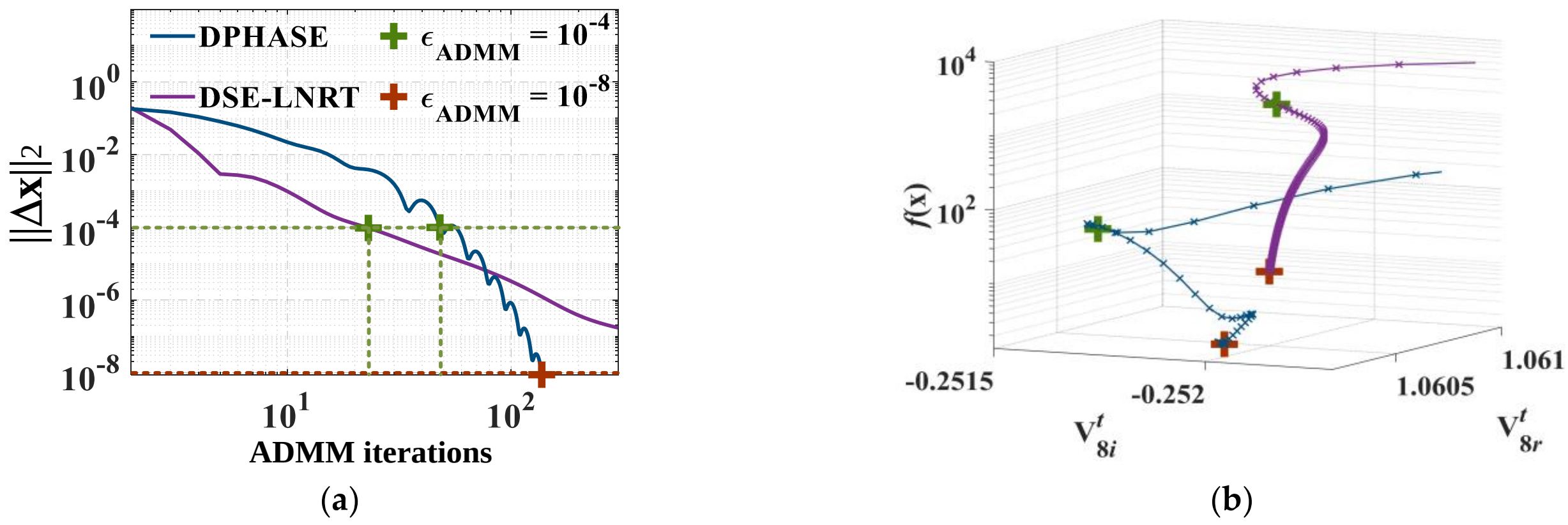

**Figure 18.** Variations (**a**) in the errors of the estimated states and (**b**) in the objective values, by the number of ADMM iterations for different convergence thresholds $\varepsilon_{ADMM}$.

**Table 11.** Comparison of the AMAEs for the corrupted SCADA data.

| AMAE [×10⁻³] | Proposed | | Conventional | | | | | |
|---|---|---|---|---|---|---|---|---|
| | DPHASE | | DSE *w/o* BDP | | DSE-LNRT | | RDSE | |
| | $\lvert V \rvert$ | $\theta$ | $\lvert V \rvert$ | $\theta$ | $\lvert V \rvert$ | $\theta$ | $\lvert V \rvert$ | $\theta$ |
| 14-bus | 0.05 | 0.21 | 1.04 | 1.01 | 0.75 | 0.69 | 0.62 | 0.69 |
| 118-bus | 0.19 | 0.24 | 3.84 | 4.02 | 1.75 | 2.92 | 1.99 | 3.52 |
| 1062-bus | 0.17 | 0.18 | 8.99 | 9.57 | 3.08 | 3.23 | 4.37 | 8.52 |

**Table 12.** Comparison of the AMAEs for the corrupted PMU data.

| AMAE [×10⁻³] | Proposed | | Conventional | | | | | |
|---|---|---|---|---|---|---|---|---|
| | DPHASE | | DSE *w/o* BDP | | DSE-LNRT | | RDSE | |
| | $\lvert V \rvert$ | $\theta$ | $\lvert V \rvert$ | $\theta$ | $\lvert V \rvert$ | $\theta$ | $\lvert V \rvert$ | $\theta$ |
| 14-bus | 1.20 | 1.97 | 66.8 | 27.9 | 55.4 | 22.5 | 65.5 | 25.3 |
| 118-bus | 7.2 | 5.9 | 84.5 | 56.8 | 63.3 | 46.6 | 77.0 | 54.7 |
| 1062-bus | 13.1 | 14.5 | 126.6 | 93.3 | 107.8 | 92.4 | 108.6 | 91.4 |

**Table 13.** Buses in the five sub-areas in the IEEE 118-bus network.

| Sub-Areas | Buses † |
|---|---|
| 1 | 1, 2, 3, 4, 5, 6, 7, 8, 9, 10, 11, 12, 13, 14, 15, 16, 17, 18, 19, 20, 117 |
| 2 | 21, 22, 23, 24, 25, 26, 27, 28, 29, 30, 31, 32, 70, 71, 72, 73, 74, 75, 113, 114, 115 |
| 3 | 33, 34, 35, 36, 37, 38, 39, 40, 41, 42, 43, 44, 45, 46, 47, 48, 49, 50, 51, 52, 53, 54, 55, 56, 57, 58, 59 |
| 4 | 68, 69, 76, 77, 78, 79, 80, 81, 82, 83, 84, 85, 86, 87, 88, 89, 90, 91, 92, 93, 94, 95, 96, 97, 98, 99, 116, 118 |
| 5 | 60, 61, 62, 63, 64, 65, 66, 67, 100, 101, 102, 103, 104, 105, 106, 107, 108, 109, 110, 111, 112 |

(† The underlined numbers indicate the buses where the PMUs are located).

## 4. Additional Research

To realize 100% renewable electricity, the generation and transmission capacity must deal with all VRE issues, including weather dependence and power electronics interfaces, in a reliable and cost-effective manner. Moreover, electricity market design and clearing should include long-term incentives and investments toward an optimal energy mix and a market-driven capacity expansion, given a trade-off between reliability and cost.

### *4.1. Energy and Capacity Adequacy*

Attainment of a zero-carbon grid requires careful planning. Power plants and transmission facilities must meet all the demand at all locations and at all times. In general, energy adequacy planners were traditionally of concern mainly for grids dominated by hydroelectricity but will become ever more important for VRE-dominated grids. The planners commonly explore the metrics for the loss of load probability (LOLP) or the loss of load expectation (LOLE) [80]. Moreover, even if power-plant numbers are adequate, transmission capacity may be lacking and if so, traditional planners built more peak-load plants. However, other options become more relevant as VRE proportions increase. These include contributions from behind-the-meter storage in forms of electrical and other energy within the same and neighboring regions. Assessment of capacity adequacy relies on the capture of rare risks, for example, when a high load demand triggers simultaneous generation/transmission failure [81].

#### 4.1.1. Adequacy Planning Tools

New planning tools must accurately consider the stability, reliability, and flexibility of grid operation [82]. Most current planning tools seek to optimize the generation mix by solving a cost minimization problem, while assessing generation and transmission adequacy. Given the scenarios of transmission capacity expansion, the problem aims to only minimize the total investment cost that optimizes generation capacity by reference to the relevant technologies and their costs [83]. The ability to detail investment for transmission capacity expansion will then become limited as the VRE proportion approaches 100% [84]. Moreover, to render the problem tractable, the current planning tools employ rather simplified representations of generator details. The tools do not consider all relevant aspects of generator operation and, thus, the path toward 100% renewable electricity should be demonstrated using iteration between planning and operation, thus multi-year time-series scheduling at high temporal resolution. Another challenge is the increasing need for energy sector coupling, thus maximizing demand flexibility in, and the welfare of, the entire energy system by incorporating electricity demand with other energy sectors. A further need is how energy sector coupling can ensure storage on the seasonal time scale, assisting to resolve the energy and capacity adequacy issue. Adapted and enhanced tools are needed that incorporate new constraints on existing planning models or the ability to link planning models with more detailed analytical tools. This is because the traditional tools do not consider the impacts of important enablers of the zero-carbon power grid: transmission expansion, distributed generation, energy storage, demand participation, and sector coupling.

#### 4.1.2. Metrics and Calculation Methodologies

In a 100% renewable electricity system, the bulk of investment is a capital expense, primarily driven by energy requirements and the adequacy of capacity [85,86]. VRE sources contribute to both of these. However, their contribution decreases as the VRE proportion increases, as for most types of energy. The decreases are marked mainly because the characteristics of VRE sources are inherently correlated but can thus be mitigated via the diversification and creation of large geographical footprints. In other words, grid operators should invest in diverse and less-correlated VRE sources [87], and in energy storage and demand-side participation, thus ensuring the adequate energy and capacity without overinvestment. Note that demand-side participation does affect the adequacy, although in a rather unidirectional manner; electricity is transformed into another form of energy that is not re-transformed back to electricity. Consequently, as the VRE proportion rises to 100%, grid operators must re-examine their classical adequacy metrics (e.g., the loss of load and the expected unserved energy) and update adequacy calculation methodologies; metrics and methods that consider wind and solar PV are rather well developed but not widely applied in practice.

Classical metrics are inappropriate when energy storage systems are adequate and the loads are both flexible and responsive. Given more flexible resources and more demand-side participation in a zero-carbon power grid, the adequacy metrics need to be updated to reflect the needs regarding critical loads, consumer types, and network sizes, for example [88]. The LOLP and LOLE are rather arbitrarily employed without consideration of the requirements, and the expected unserved energy (EUE) may be more meaningful if energy limitations become more dominant than capacity limitations. The LOLP and EUE are useful if a large proportion of demand is totally inflexible—i.e., demand must be served. If enough demand is flexible and responsive, the classical metrics recede and should be replaced by new metrics. There is also a need for more rigorous methods of adequacy calculation; adequacy should be ensured at a minimal cost and rewarded in a market setting (or otherwise) without overinvestment. The classical adequacy problem is then replaced by a cost minimization problem, wherein the grid investment cost is balanced with the cost of energy storage system installation and reductions (or shifts) in consumer load demands.

#### 4.1.3. Contributions of Emerging Technologies

The assessment of the energy and capacity adequacy needs to reflect the costs for demand response, energy storage, distributed generation, and sector coupling, in addition to the cost for transmission line expansion. Specifically, most existing adequacy metrics assume fixed loads. Thus, any load reduces system adequacy. This means that consumer (load) types should be accurately assessed. Some load types (EVs and HVAC units) are untapped resources. The energy need is relatively constant, but the times of battery charging and heating- and cooling-energy use are at least somewhat flexible. The trade-off between cost and reliability means seeking a reliability level that considers cost efficiency [2]. It also would include customer disconnection during hours of scarcity, thus rolling blackouts when the load responsiveness to price is inadequate. Moreover, storage devices should be better represented in the metrics, with the incorporation of state-of-charge limitations during hours of scarcity. The operational characteristics of storage devices, if deployed at a large scale, should be better modeled in terms of their contributions to adequacy. Any contribution of distributed generation to adequacy can be limited by the distribution networks and their operational characteristics, which depend on weather-associated parameters, such as temperature, wind speed, and solar irradiance. The contributions of sector coupling across transportation, heating, cooling, gas, and hydrogen are currently difficult to assess, highlighting the need for improvements of current planning tools.

#### 4.1.4. A List of Research Challenges

Table 14 lists the challenges in estimating energy and capacity adequacy within desired confidence levels. The dimensionality and underlying correlations between supply and demand indicate that higher-quality models and longer time-frame data are required to render adequacy calculations that are both accurate and robust. In essence, the modeling and data needs are those relevant to the effective assessment of the potential contributions of demand response, energy storage, distributed generation, and sector coupling.

**Table 14.** List of research challenges in energy and capacity adequacy.

| Challenges | Research Questions and Requirements |
|---|---|
| Adequacy planning tools | • ***Inclusion of short-term supply-and-demand balancing when exploring the effects of forecast uncertainty*** on the optimal VRE mix. |
| Metrics and calculation methodologies | • ***Use of several current metrics,*** including LOLP, LOLE, and loss-of-load hours (LOLH), and ***development of new metrics*** that consider societal needs and load and storage flexibilities.<br>• ***Use of different reliability levels,*** such as an event occurring once in 10 years, as the common LOLP target and two events per year as a lower reliability target.<br>• ***Improving the calculation of adequacy benefits afforded when neighboring regions are interfaced and larger geographical areas are connected*** via transmission lines with limited capacities. |
| Contributions of emerging technologies | • ***Improving representations of demand-side flexibility, energy storage, sector coupling, grid limitation, and expansion cost,*** to better predict the investments required for grids with high VRE proportions and accurately determine the optimal VRE mix. |
| Modeling and data | • ***Transmission networks and distributed generators*** should be modeled in a manner that accurately captures their contributions to adequacy.<br>• The contribution of ***demand and storage*** to adequacy must be determined accurately using models of real-world operating characteristics, including responses to electricity prices.<br>• New models must capture how ***sector coupling among electricity, transportation, heating, and natural gas affects adequacy.***<br>• ***More high-quality longer time frame (10+ years) data on supply and demand power*** are required to ensure precise and robust calculations of adequacy. |

## 4.2. Electricity Market Design

Well-designed electricity markets play key roles in cost-effective transformations of the electric power industry while ensuring economical outcomes. Electricity markets worldwide can vary substantially in their details, driven by their energy resources and their historical, social, and political contexts. Competitive, deregulated electricity markets operate reasonably well in many parts of the United States and Europe. Other parts, such as South Korea, include regulated electricity markets only with vertically integrated utilities or adopt hybrid deregulated/regulated electricity markets. Regardless of market type, the basic principle is always the cost-effective and reliable operation of electric power grids.

### 4.2.1. Changes in Electricity Markets

In various algorithms and simulation tools, day-ahead, intra-day, and short-term market models have been widely explored to balance supply and demand and provide energy and grid services in a cost-effective manner. Given the increased installation and utilization of renewable energy resources, technological developments in the zero-carbon power grid context have greatly affected the design and operation of electricity markets, and vice versa [89–91]. For example, the replacement of traditional resources with new VRE sources is likely to remove the marginal cost of energy, significantly affecting electricity market structures and clearing mechanisms.

Recently, interest in local market designs and peer-to-peer trading of various market services has grown [92]. Many efforts are being made to link local markets with each other and to bulk markets, particularly in regions where marked expansions of distributed, behind-the-meter VRE are expected. Horizontal and vertical interconnections within the electricity market are crucial to allow large- and small-scale interplay between VRE and flexible energy resources. It is challenging to develop local- and system-level market models that accurately consider low-voltage VRE growth.

### 4.2.2. Electricity Prices and Investment Signals

Increasing VRE proportions in several markets have created increasingly longer periods of zero or negative energy prices because of excessive energy supply, including the substantial amounts of hydroelectric energy that have been generated for decades. Time-varying and inflexible energy generation can also create periods of high energy prices because energy or grid services is/are related [93]. Therefore, operators of zero-carbon power grids should be prepared for, and robust against, high price volatility. In parallel, the volume of price-responsive loads will increase because of electrification and sector coupling, leading to a paradigm shift in power commitment, dispatch, and balancing. Due to volatile energy prices, a large volume of flexible, price-responsive demand is likely to move toward periods when flexible generators are commonly available, inverting the objective of optimal operational scheduling of generators and loads. The accelerated dynamics of supply and demand are attributable to the increased uncertainty level, requiring a change from deterministic methods for risk assessment to probabilistic methods. This will assist in mitigating flexibility requirements and increasing reliability benefits. Although several methods are already capable of probabilistic assessment, the industry take-up remains low [2].

Traditionally, such price volatility encourages investment in flexible energy resources, but it is unclear whether the current market structures and clearing mechanisms will continue to produce supportive investment signals as we proceed toward 100% renewable electricity [94]. In other words, carbon neutrality can render it difficult to ensure that grid investment signals are correct; this is a perennial problem of electricity market design and operation. Some grid and market operators have begun to experiment with various energy sources and ancillary services to establish the required flexibility, and others have introduced forward markets for generation capacity to provide a revenue stream for resources that are required but as yet not paid for because of, for example, the imposition of price caps [95].

4.2.3. A list of research challenges

Table 15 summarizes the basic market challenges discussed above.

**Table 15.** List of research challenges in electricity market design.

| Challenges | Research Questions and Requirements |
|---|---|
| Changes in electricity markets | • ***Enhance market and regulatory frameworks*** to ensure that demand becomes more responsive to price in a manner that resolves potential market challenges, ***including price volatility, revenue imbalance, resource deficiency, and reduced flexibility.*** |
| Electricity prices and investment signals | • Establish well-organized markets that ***incentivize long-term investment in an optimal mix of energy resource***s ***by accurately valuing those resources and the required attributes of future electricity systems with 100% renewables.***<br>• Design market-based approaches that ***incentivize flexible use of energy resources guided by the trade-off between cost and reliability and between the valuations of various market participants.*** |

## 5. Conclusions

This paper surveyed the principal research challenges that must be addressed on the path toward 100% renewable electricity worldwide, by comprehensively searching for the recent literature, most of which were published during the period from 2016 to 2022. In particular, the previous works were selected that are directly relevant with the four main research challenges, i.e., supply-and-demand balancing, inverters as VRE interfaces, energy and capacity adequacy, and electricity market design. Two of the most important challenges are reliable, cost-effective supply-and-demand balancing and the interfacing of many VRE inverters. The selected literature was then classified into the studies in 15 sub-areas, and given the sub-area studies, the specific challenges and future research directions were outlined in each sub-area. Moreover, this paper selected the comparative case studies from the literature and discussed their results to address the main research challenges. The new algorithms and strategies for improved grid operation, estimation, and protection were explained with the numerical results from the perspectives of originality, feasibility, and effectiveness. This paper also covered the challenges posed by the energy and capacity adequacy and electricity market design; further analyses and case studies are required. We seek to assist universities, research institutes, and funders to focus and prioritize their efforts and investments, ultimately connecting all participants to resolve the diverse research challenges, including those presented, that must be overcome before reliable, economical zero-carbon power grids become the global norm.